\documentclass[aps,prb,10pt,longbibliography,superscriptaddress,twocolumn,letterpaper,nobalancelastpage,raggedbottom]{revtex4-2}

\usepackage{mycommands}

\renewcommand{\section}[1]{\textit{#1}. ---}

\begin{document}

    \newcommand{\articletitle}{Ultrastrong Coupling Signatures in Photon Statistics from Terahertz Higgs-Polaritons}
    
    \title{\articletitle}

    \author{Spenser Talkington}
    \email{spenser@upenn.edu}
    \affiliation{Department of Physics and Astronomy, University of Pennsylvania, Philadelphia, Pennsylvania 19104, USA}
    
    \author{Benjamin Kass}
    \affiliation{Department of Physics and Astronomy, University of Pennsylvania, Philadelphia, Pennsylvania 19104, USA}

    \author{Martin Claassen}
    \email{claassen@sas.upenn.edu}
    \affiliation{Department of Physics and Astronomy, University of Pennsylvania, Philadelphia, Pennsylvania 19104, USA}
    
    \date{\today}
    
    \begin{abstract}
        The ultrastrong coupling regime of cavity photons and quantum materials  has emerged as a pathway to modify materials properties, however definitive signatures of ultrastrong coupling remain elusive. Focusing on the quantum photon statistics of light transmitted through a cavity-embedded superconductor, we show that a two-photon Higgs polariton at strong coupling realizes a photonic nonlinearity at the single terahertz photon level. We find that as light-matter coupling increases, the photon statistics show pronounced changes due to the formation of a hybrid photon-matter dark-cavity state with finite photon occupancy, producing testable signatures of ultrastrong coupling. We derive a non-Markovian input output relation and study the cavity-embedded superconductor 2H-NbSe$_2$ as it approaches ultrastrong light-matter coupling. Our results reveal a diagnostic for ultrastrong coupling in the two-photon coincidence statistics that is absent in total counts.
    \end{abstract}
    
    \maketitle
    
    \section{Introduction}\label{sec:intro}
        Ultrastrong light-matter coupling in cavities, where vacuum fluctuations of the confined electromagnetic field can modify the ground state of a quantum material, has emerged as a frontier for manipulating material properties \cite{schlawin2022cavity,bloch2022strongly,hubener2024quantum,garcia2021manipulating,bretscher2026fluctuation}, including quantum Hall states \cite{appugliese2022breakdown,enkner2025tunable,graziotto2025cavity}, superconductivity \cite{sentef2018cavity,schlawin2019cavity,curtis2019cavity,thomas2025exploring,keren2026cavity,xu2026vacuum}, metal-insulator transitions \cite{jarc2023cavity,fassioli2025controlling}, polaritons \cite{basov2020polariton,smolka2014cavity,maissen2014ultrastrong,paravicini2019magneto,keller2020landau}, quantum criticality \cite{weber2023cavity,kass2024many,sur2025amplified}, and more \cite{ashida2021cavity,chiocchetta2021cavity,vinas2023controlling,eckhardt2022quantum,passetti2023cavity,shaffer2024entanglement,kiffner2019manipulating,jarc2023cavity,nambiar2025diagnosing,grunwald2025cavity}. THz cavities are especially promising, as the photon energy can be resonant with collective modes of quantum materials. In such systems, the photon-matter coupling strength can exceed the cavity linewidth (``strong coupling'') and approach or surpass the bare cavity frequency \cite{kim2025symmetry,baydin2025terahertz} (``ultrastrong'' or ``deep strong'' coupling). In this regime, the dark-cavity ground state acquires hybrid photon-matter properties, which can drastically alter macroscopic properties. However, direct quantum signatures of hybrid photon-matter phases at ultrastrong coupling remain elusive.
        
        The Higgs (amplitude) mode of a cavity-embedded superconductor is a particularly promising platform for observing such quantum signatures from the hybridization of THz photons and superconducting order parameter. Single-photon excitations of Higgs modes are symmetry forbidden by gauge invariance; instead, they couple to photon pairs \cite{schwarz2020classification}. At ultrastrong coupling, where the two-photon excitation amplitude exceeds the cavity linewidth, a Higgs two-photon polariton can form. This scenario differs from single-photon Higgs polaritons in driven systems that require assistance via super-currents \cite{moor2017amplitude,nakamura2019infrared,allocca2019cavity,raines2020cavity} or population inversion \cite{glier2025non}.
        The avoided two-photon Higgs resonance can realize a THz photon blockade -- far from conventional optical or microwave frequency schemes in atomic, quantum dot, exciton-polariton, and circuit and waveguide QED systems \cite{ridolfo2012photon,flayac2013input,muller2015coherent,goto2019figure,kass2024many,heinisch2026high,trivedi2019photon,chen2022photon,delteil2019towards,sheremet2023} -- with immediate consequences for the photon statistics of transmitted light through the cavity. Closely related nonlinear few-photon processes have been studied in waveguide QED, where localized emitters mediate scattering and generate nonclassical correlations between photons \cite{sheremet2023}. Moreover, and central to this work, reaching ultrastrong coupling with dark-cavity photon-matter states with finite photon occupation leads to drastic deviations from the simple blockade scenario.
    
        In this work, we show that photon statistics and the second-order photon coherence $g^{(2)}$ can serve as a witness of ultrastrong coupling of Higgs two-photon polaritons in cavity superconductors and realize a THz photon blockade regime. We focus on the superconductor 2H-NbSe$_2$ and study the second-order photon coherence
        \begin{align}
            g^{(2)}(\tau) = \frac{\langle \hat{b}_\mathrm{out}^\dagger(t)\hat{b}_\mathrm{out}^\dagger(t+\tau)\hat{b}_\mathrm{out}(t+\tau)\hat{b}_\mathrm{out}(t)\rangle}{\langle \hat{b}_\mathrm{out}^\dagger(t+\tau)\hat{b}_\mathrm{out}(t+\tau)\rangle\langle \hat{b}_\mathrm{out}^\dagger(t)\hat{b}_\mathrm{out}(t)\rangle}
        \end{align}
        where $\hat{b}_{\mathrm{out}}(t) = \int d\omega \ e^{-i\omega(t-t_f)} \hat{b}(\omega)$  is the output photon annihilation operator for a final reference time $t_f$. In transmission geometry with a weak classical THz drive, equal time coincidence measurements $g^{(2)}(0)$ \cite{mandel1959fluctuations,glauber1963quantum,brown1956correlation,paul1982photon,chang2014quantum} diagnose photon bunching and antibunching. We first study the weak and intermediate coupling regime with a trivial cavity ground state analytically, then develop a non-Markovian input-output framework to diagnose photon statistics at ultrastrong coupling, necessitated by the breakdown of the Markov approximation for finite dark-cavity photon fluctuations \cite{shen2005coherent,shen2007strongly,fan2010input,ciuti2005quantum,ciuti2006input,lednev2024lindblad,kim2026photon}. By comparing these results, we identify signatures of ultrastrong coupling that are invisible in transmission measurements. We analyze the results for 2H-NbSe$_2$ and discuss generalizations to other materials platforms.

    \section{Model}
\begin{figure}
            \centering
            \includegraphics[width=\linewidth]{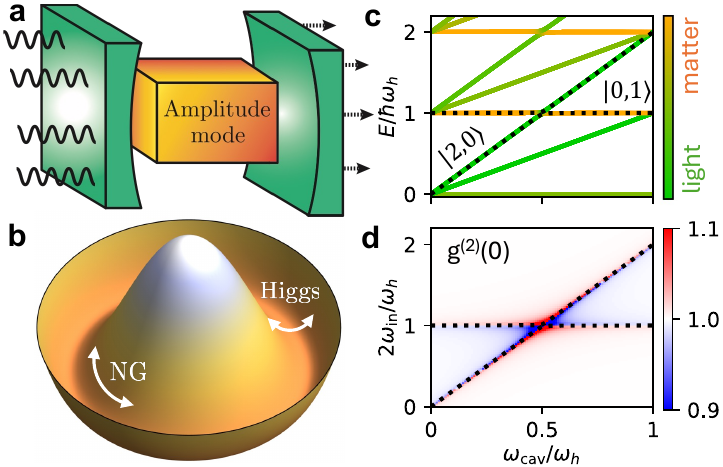}
            \caption{THz photon blockade from cavity two-photon Higgs polaritons. \textbf{(a)}, \textbf{(b)} Cavity-embedded superconductors with Higgs (amplitude) modes -- fluctuations in the superconducting order parameter about its equilibrium state -- can realize two-photon nonlinearities, witnessed via transmitted photons in response to a weak classical drive. The Nambu-Goldstone mode (NG) is pushed to the plasma frequency by the Anderson-Higgs mechanism. \textbf{(c)} Spectrum of the Higgs-photon Hamiltonian as a function of cavity frequency with up to two quanta in each mode $|n_\mathrm{cav},n_h\rangle$. Colors show $\langle n_h\rangle/(\langle n_\mathrm{cav}\rangle+\langle n_h\rangle)$. Dotted lines indicate the  Higgs two-photon polariton branches. \textbf{(d)} Strong-coupling photon blockade in the Higgs polariton gap leads to anti-bunching $g^{(2)}(0) \to 0$, plotted for  weak driving analytic results from Eq. \ref{eq:g2-resonance} with  $\kappa/\hbar\omega_h=0.01$, $\gamma/\hbar\omega_h=0.002$, and $\gamma_h/\hbar\omega_h=0.04$.}
            \label{fig:fig1}
        \end{figure}

        We start from a generic material with a collective mode, such as the Higgs mode, embedded in a THz cavity illuminated with light in transmission geometry. We illustrate the proposed setup in Fig. \ref{fig:fig1}(a).
        Consider a superconductor integrated with a cavity that hosts a single photon mode $\hat{H}_\mathrm{cav}=\hbar \omega_\mathrm{cav} \hat{a}^\dagger \hat{a}$ with a frequency $\omega_\mathrm{cav}$ below the superconducting gap $\Delta$. The elementary collective matter excitation in this regime is the Higgs mode, corresponding to fluctuations of the superconducting order parameter around the condensate minimum. The Higgs mode frequency coincides with pair-breaking excitations $2\Delta$ at mean field level, but is pushed to lower energies by strong-coupling effects or hybridization with other modes. Single-photon excitation is symmetry-forbidden; instead, the Higgs mode couples to photon pairs via a diamagnetic $A^2$ coupling. Starting from a Ginzburg-Landau free energy for the superconducting gap order parameter $\Delta$ \footnote{See Supplemental Material Section \ref{sec:TDGL} for the Ginzburg-Landau coefficients in terms of physical quantities.} 
        \begin{align}\label{eq:free}
            F = - \alpha |\Delta|^2 + \beta|\Delta|^4 + K|(i \nabla + 2eA/\hbar)\Delta|^2
        \end{align}
        with $\alpha,\beta > 0$ and gradient $K$, an effective action for amplitude oscillations can be derived by expanding $\Delta = (\Delta_0 + \Delta')e^{i\theta}$ in amplitude fluctuations $\Delta'$ around the minimum $\Delta_0 = \sqrt{\alpha/2\beta}$ \cite{shimano2020higgs,pekker2015amplitude}. Quantizing the Higgs mode $\Delta' \to \Delta_\mathrm{zpf} (\hat{h} + \hat{h}^\dagger)$ and electromagnetic field $A \to A_\mathrm{zpf} (\hat{a} + \hat{a}^\dagger)$ with $A_\mathrm{zpf}=\sqrt{\hbar/2\omega_\mathrm{cav}\epsilon_0 V_\mathrm{EM}}$, mode volume $V_\mathrm{EM}$ and $\Delta_\mathrm{zpf} = 1/2\sqrt{2\Delta_0 PV_\mathrm{material}}$ (where $P$ is coefficient of the inertial coupling $|\hbar\partial_t \Delta|^2$ of the time-dependent Ginzburg-Landau action), gives an effective Hamiltonian for Higgs-photon interactions
        \begin{align}\label{eq:H}
            \hat{H} = \hbar \omega_h \hat{h}^\dagger \hat{h} + \hbar \omega_\mathrm{cav} \hat{a}^\dagger \hat{a} + \kappa(\hat{a}+\hat{a}^\dag)^2(\hat{h}+\hat{h}^\dag) ~.
        \end{align} 
        Here, $\hbar\omega_h=\sqrt{2\alpha/P}$ and $\kappa=K (2e)^2 A_\mathrm{zpf}^2 2\Delta_0 \Delta_\mathrm{zpf}/\hbar^2$ is the intrinsically \textit{nonlinear} light-matter coupling strength. To stabilize this cubic Hamiltonian, we include a small quartic term $\kappa_4(\hat{a}+\hat{a}^\dag)^4$ with $\kappa_4=\kappa^2/\hbar\omega_h$. This term ensures the stability of the perturbative fluctuation expansion around the superconducting state, and such cavity coupling to a Raman-active mode was shown to enhance dark-cavity photon fluctuations \cite{ojeda2024equilibrium}. When coherently excited by many photons $\hat{H}$ causes parametric amplification as seen with K$_3$C$_{60}$ Higgs modes \cite{buzzi2021higgs}. Near the two-photon resonance with the Higgs mode $2 \omega_\mathrm{cav} = \omega_h$, a hybrid excited state of two photons and one Higgs excitation forms: a \textit{two-photon} Higgs polariton. In stark contrast to optical modes in atomic-molecular-optical (AMO) settings, THz cavities routinely access the ultrastrong regime \cite{kim2025symmetry,baydin2025terahertz}, where the dark cavity state contains photon excitations. Detecting signatures of this regime remains challenging as standard measurements of polaritonic splittings are largely insensitive to the restructured ground state \cite{forn2019ultrastrong,frisk2019ultrastrong}. As we show below, $g^{(2)}$ provides a diagnostic -- revealing qualitative departures from weak coupling that are invisible in linear response.

    \section{Conventional Strong Light-Matter Coupling}
        \begin{figure*}
            \centering
            \includegraphics[width=\linewidth]{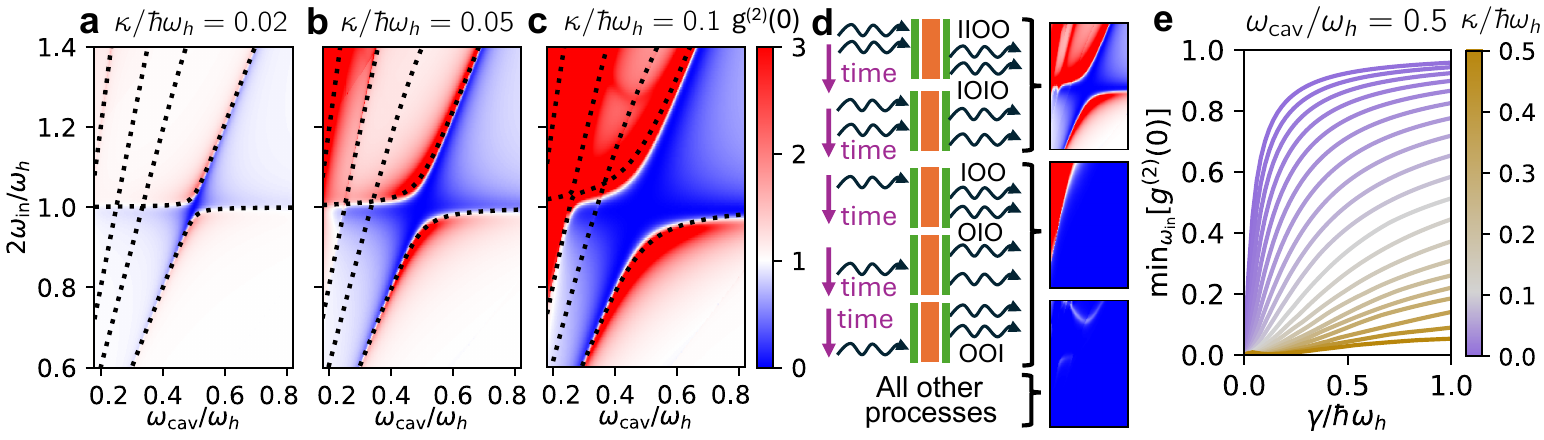}
            \caption{
            Photon statistics of hybrid photon-matter states at ultrastrong coupling. \textbf{(a)}, \textbf{(b)}, \textbf{(c)} depict $g^{(2)}(0)$ vs input and cavity frequency, upon approaching ultrastrong coupling for $\kappa/\hbar\omega_h=0.02, 0.05, 0.1$, respectively. Dashed lines indicate lower/upper two-photon polaritons and states with large support on $|n,0\rangle$ for $n=3,4,5$ photons. While these deviate weakly from RWA, the photon statistics shows striking differences, rooted in a hybrid photon-matter dark-cavity ground state. Parameters are $\gamma/\hbar\omega_h=0.005$, $\gamma_h/\hbar\omega_h=0.02$. \textbf{(d)} Decomposition of $g^{(2)}(0)$ for (c) into scattering processes, depicted schematically. The $IIOO$ and $IOIO$ processes capture anti-bunching from the avoided two-photon resonance in the Higgs polariton gap; additional higher polariton resonances due to finite dark-cavity photon occupancy contribute bunching. At ultrastrong coupling, stimulated emission due to the $IOO$, $OIO$, $OOI$ processes emerges, capturing the ejection of two dark-cavity photons upon the injection of one input photon. Further scattering processes remain small below $\kappa / \hbar\omega_h \sim 0.5$.  \textbf{(e)} Minimum of $g^{(2)}(0)$ for $2\omega_\mathrm{cav} = \omega_h$ upon sweeping the input frequency $\omega_\mathrm{in}$. High-quality cavities and stronger light-matter coupling favor antibunching. The stimulated emission and higher polariton features are invisible in transmission but are sensitive to the onset of ultrastrong coupling in $g^{(2)}$.}
            \label{fig:fig2}
        \end{figure*}
        We first analyze the case of intermediate coupling where $\kappa$ exceeds the photon linewidth but satisfies $\kappa \ll \omega_\mathrm{cav}$. In this case, the dark-cavity ground state is a trivial vacuum state. Near the two-photon resonance, a rotating wave approximation (RWA) retains only resonant couplings $\hat{H}_\mathrm{RWA} = \hbar\omega_\mathrm{cav} \hat{a}^\dagger \hat{a} + \hbar\omega_h \hat{h}^\dagger \hat{h} + \kappa (\hat{h}^\dagger \hat{a} \hat{a} + \hat{h}\hat{a}^\dag\hat{a}^\dag)$ and permits an analytic solution to the steady state output photon statistics for a coherent THz input.
        The two-photon upper and lower Higgs polariton branches at weak coupling are depicted in Fig. \ref{fig:fig1}(c) and are obtained from
        \begin{align}\label{eq:polariton}
            H_p = \begin{bmatrix}
            \langle 2,0|\hat{H}|2,0\rangle\! & \langle 2,0|\hat{H}|0,1\rangle\\
            \langle 0,1|\hat{H}|2,0\rangle\! & \langle 0,1|\hat{H}|0,1\rangle
            \end{bmatrix}
            = \begin{bmatrix}
            2\hbar\omega_\mathrm{cav} & \sqrt{2}\kappa\\
            \sqrt{2}\kappa & \hbar\omega_h
            \end{bmatrix}
        \end{align}
        where $|n_\mathrm{cav},n_h\rangle$ denotes the many-body Fock space. Importantly, in contrast to conventional polaritons where the one-photon state hybridizes with matter, here the one-photon state is unaffected but the two-photon state opens a polaritonic gap due to hybridization with the Higgs mode. This has immediate consequences for the output photon statistics. Suppose that an input field is applied at $\omega_\mathrm{cav}$, permitting resonant tunneling of single photons: tunneling a second photon into the cavity is off-resonant due to the two-photon polaritonic gap. If the polariton linewidth is less than $\kappa$, two-photon transmission becomes blockaded and $g^{(2)}(0)\to 0$ is antibunched.
        
        To compute $g^{(2)}(0)$ in RWA, we introduce a coherent driving $\hat{H}_\mathrm{in} = f \hat{a}^\dag + \bar{f}\hat{a} - \hbar\omega_\mathrm{in}\hat{a}^\dag\hat{a} - 2\hbar\omega_\mathrm{in}\hat{h}^\dag\hat{h}$
        and solve for the steady state of the Lindblad master equation \cite{stefanini2025lindblad}:
        \begin{align}
            \hbar\partial_t \hat{\rho} = -i[\hat{H}_\mathrm{RWA} \!+\! \hat{H}_\mathrm{in},\hat{\rho}] +\! \gamma\, \mathcal{D}[\hat{a},\hat{\rho}] +\! \frac{\gamma_h}{2} \mathcal{D}[\hat{h},\hat{\rho}]
        \end{align}
        with $\mathcal{D}[\hat{J},\hat{\rho}]=2\hat{J}\hat{\rho} \hat{J}^\dagger - \hat{J}^\dagger \hat{J}\hat{\rho} - \hat{\rho} \hat{J}^\dagger \hat{J}$ and decay rates $2\gamma$, $\gamma_h$ for the (two-sided) cavity and Higgs modes. The exact steady state of this master equation can be obtained in the $\gamma_h\gg 2\gamma$ limit \cite{drummond1980non}. In RWA, correlation functions of the output field $\hat{b}_\mathrm{out}(t)$ are related to intra-cavity photon correlation functions via the input-output relation in Markov approximation $\hat{b}_\mathrm{out}(t) = \hat{b}_\mathrm{in}(t) - i\sqrt{\gamma} \hat{a}(t)$ \cite{gardiner1985input,clerk2010introduction,carusotto2013quantum}. Assuming a vacuum input state on the detector side, we obtain $g^{(2)}(0)=(I_{22}/I_{00})/(I_{11}/I_{00})^2$ with \footnote{See Supplemental Material Section \ref{sec:analytic} for derivation of correlation functions from the Markovian master equation.}
        \begin{align}\label{eq:analytic}
        I_{nn} = |\phi|^{2n} \bigg|\frac{\Gamma(j)}{\Gamma(j+n)}\bigg|^2 {_0}F_2(j+n,\bar{j}+n;2|\phi|^2)
        \end{align}
        for hypergeometric function ${_0}F_2$. Detunings are $\delta_\mathrm{cav}=\omega_\mathrm{cav}-\omega_\mathrm{in}$, $\delta_h=\omega_h-2\omega_\mathrm{in}$, and $j = (\gamma+i\delta_\mathrm{cav})(\gamma_h/2+i\delta_h)/\kappa^2$,  and $\phi=-if(\gamma_h/2+i\delta_h)/\kappa^2$.
        At weak drive
        \begin{align}\label{eq:g2-resonance}
            g^{(2)}(0) = \left|\frac{(\gamma+i\delta_\mathrm{cav})(\gamma_h/2+i\delta_h)}{\kappa^2 + (\gamma+i\delta_\mathrm{cav})(\gamma_h/2+i\delta_h)}\right|^2
        \end{align}
        confirming that $g^{(2)}(0) \to 0$ at the center of the polaritonic gap for sufficiently high cavity quality factors.
        This is mirrored in Fig. \ref{fig:fig1}(d) where bunching (from a two-photon resonance with the upper or lower polariton) gives way to anti-bunching in the middle of the polaritonic gap.

    \section{Ultrastrong Light-Matter Coupling}
        A hallmark of ultrastrong coupling is a dark-cavity \textit{ground} state with hybrid matter and photon characteristics. As $\kappa$ approaches $\omega_\mathrm{cav}$ and the RWA fails, the ground state acquires a weak but finite dark occupation and is \footnote{See Supplemental Material Section \ref{sec:GS-fock} for an explicit evaluation of the coefficients $c_{N_\mathrm{cav},0}$.}
        \begin{align}\label{eq:gs}
        |0\rangle \!= c_{0,0}|0,\!0\rangle + c_{2,0}|2,\!0\rangle + c_{0,1}|0,\!1\rangle + c_{4,0}|4,\!0\rangle + \dots
        \end{align}
        which is an even photon-number parity eigenstate that contains virtual photon pairs which enable new scattering pathways for photon transmission such as stimulated emission of dark-cavity photons and coupling to higher polariton branches. These processes are rendered visible by measuring the output photon statistics, which serves as a key signature for reaching ultrastrong coupling.

        \begin{figure*}
            \centering
            \includegraphics[width=\linewidth]{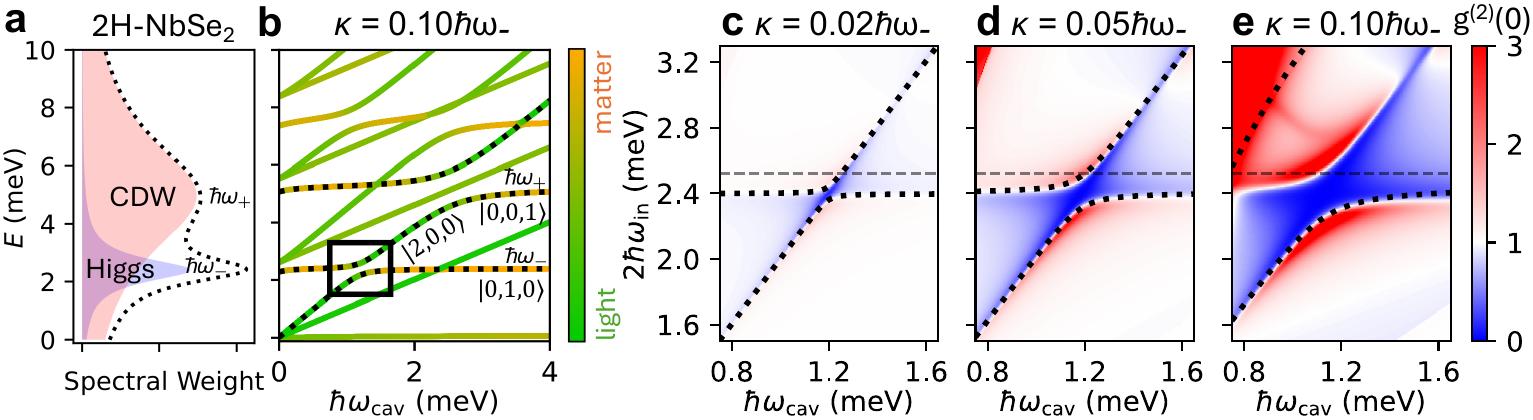}
            \caption{Frequency dependent photon statistics of cavity-embedded superconducting 2H-NbSe$_2$. \textbf{(a)} 2H-NbSe$_2$ features two collective modes at $\hbar\omega_-=2.40$ (mostly Higgs) and $\hbar\omega_+=4.96$ meV (mostly CDW). We fix $\hbar\omega_h=2.52$ meV, $\hbar\omega_c=4.84$, $\kappa_{hc}=0.55$, $\kappa_h=\kappa$, $\kappa_c=\kappa/0.58$, $\gamma=0.01$, $\gamma_h=0.1$, and $\gamma_c=0.2$ meV, which corresponds to $Q=120$ at $\hbar\omega_\mathrm{cav}=1.2$ meV. \textbf{(b)} As in the single mode case, anti-bunching occurs in the polaritonic gap, illustrated for the same $\kappa$ as in (e); we will focus on frequencies near the polaritonic gap denoted by the black rectangle; states are $|n_\mathrm{cav},n_h,n_c\rangle$. \textbf{(c-e)} Photon statistics near Higgs polariton crossing; $2\Delta$ is shown with the dashed gray line. Asymmetry in the $g^{(2)}$ around the polaritonic gap indicates ultrastrong coupling, and additional antibunching above the diagonal is due to the second collective mode.}
            \label{fig:fig3}
        \end{figure*}

        At ultrastrong coupling, the standard Markov approximation of input-output theory ceases to apply, as a spectrally flat photon bath extending to negative frequencies would unphysically permit virtual photons of the dark-cavity ground state to leak out \cite{ciuti2005quantum,ciuti2006input,lednev2024lindblad}. This is cured by keeping only positive bath frequencies, which leads to a \textit{non-Markovian} input-output relation $\hat{b}_\mathrm{out}(t) = \hat{b}_\mathrm{in}(t) - i \int dt'\, \Gamma(t\!-\!t') \hat{a}(t')$ where $\Gamma(\tau) = \int d\omega\, \sqrt{\gamma} \,\theta(\omega) e^{-i\omega \tau}$ is the positive-frequency Fourier transform of the tunneling amplitude $\sqrt{\gamma}$ \footnote{See Supplemental Material Section \ref{sec:io-theory} for a derivation of the non-Markovian input-output relations.}. 
        We develop a scattering-matrix formalism that relates the detected output correlations $g^{(2)}(t) = G^{(2)}(t)/[G^{(1)}(t)]^2$ to intra-cavity multi-point functions at leading order in photon tunneling. At coincidence $t=0$ \footnote{See Supplemental Material Section \ref{sec:transmission} for the transmission $G^{(1)}(0)$, and Section \ref{sec:scattering} for an analysis of scattering pathways.}\footnote{In Section \ref{sec:scattering}E we derive the conditions under which Eq. \ref{eq:g2-usc} reduces to Eq. \ref{eq:g2-resonance}.}
        \begin{align}\label{eq:g2-usc}
            G^{(1)}(0) &= \sum_{f} \bigg| \int\limits_0^\infty \frac{d\Omega \gamma}{2\pi} \frac{e^{i\Omega t_f} \bra{f}  \hat{a} \hat{\mathcal{G}}(\omega_\mathrm{in}) \hat{a}^\dagger + \hat{a}^\dagger \hat{\mathcal{G}}(-\Omega) \hat{a} \ket{0}}{E_0 - E_f + \omega_\mathrm{in} - \Omega + i\eta}  \bigg|^2 \notag\\
            G^{(2)}(0) &=\! \sum_{f} \bigg|\! \int\limits_0^\infty d\Omega \sum_j \frac{e^{i\Omega t_f} \int_0^\Omega d\omega~ N^{(f)}_{j}(\omega,\Omega)}{E_0 \!-\! E_f \!+\! n_\mathrm{in} \omega_\mathrm{in} \!-\! \Omega \!+\! i\eta} \bigg|^2
        \end{align}
        Here, $E_0, E_f$ are ground and final state energies, $\omega_\mathrm{in}$ is the frequency of the left-side input field,  $\hat{\mathcal{G}}(\omega) = ( \hbar\omega + E_0 - \hat{H}_\mathrm{cav} + i\eta )^{-1}$ is the resolvent operator for the cavity Hamiltonian where $\hat{H}_\mathrm{cav}$ includes a phenomenological broadening $\hat\Sigma=-i\gamma \hat a^\dag \hat a-i \gamma_h \hat h^\dag \hat h/2$, and $t_f \to \infty$ is the final time (which simplifies the integral over $\Omega$ to a contour integral over the positive upper quadrant of the complex plane). Single photon transmission $G^{(1)}$ entails injection of a left-side input photon followed by ejection of an output photon to the detector (denoted ``$IO$''), or ejection of an output photon from the dark-cavity ground state, followed by injection of an input photon (denoted ``$OI$''). Two-photon transmission $G^{(2)}$ is enabled by 9 processes $OOI$, $OIO$, $IOO$, $OOII$, $OIOI$, $OIIO$, $IOOI$, $IOIO$, $IIOO$, described by scattering amplitudes $N^{(f)}_{j}$ where $j=1,\dots,9$. Only $IIOO$ and $IOIO$ remain in the RWA case of zero dark-cavity occupation. Importantly, Eq. \ref{eq:g2-usc} recovers Eq. \ref{eq:g2-resonance} at small $\kappa$ \footnote{See Supplemental Material Section \ref{section:usc-to-sc} for an illustration of Eq. \ref{eq:g2-usc} recovering Eq. \ref{eq:g2-resonance}.}. For example, the $IIOO$ amplitude is constructed by replacing each $O$ with $\hat{a}$ and each $I$ with $\hat{a}^\dagger$, weighting each tunneling event by $\sqrt{\gamma/2\pi}$, and inserting resolvents between events
        \begin{align*}
            N^{(f)}_{IIOO} = \big(\frac{\gamma}{2\pi}\big)^2 \langle f| \hat{a} \hat{\mathcal{G}}(2\omega_\mathrm{in}-\omega) \hat{a} \hat{\mathcal{G}}(2\omega_\mathrm{in})\hat{a}^\dag \hat{\mathcal{G}}(\omega_\mathrm{in}) \hat{a}^\dag |0\rangle \label{eq:N}
        \end{align*}

        Fig. \ref{fig:fig2} depicts $g^{(2)}(0)$ as a function of light-matter coupling strength. Weak coupling (b) recovers the RWA results [Fig. \ref{fig:fig1}(c)] near the two-photon resonance. However, new features (discussed below) arise in the photon statistics at ultrastrong coupling [Fig. \ref{fig:fig2}(b), (c)], which originate from the nontrivial cavity dark state. The $IIOO$ process accounts for bunching and antibunching near the polaritonic anticrossing: a two-photon resonance or a two-photon blockade if $2\omega_\mathrm{in}$ is resonant with the upper/lower polariton branch or lies in the gap respectively. Meanwhile $IOIO$ produces the sharp resonances in the top panel of Fig.~\ref{fig:fig2}(d), corresponding to transitions to higher states with support on $|3,0\rangle$, $|4,0\rangle$, and $|5,0\rangle$.
        
        The non-trivial dark-cavity ground state photon occupation enables a second set of \textit{stimulated} emission ($IOO$) processes which lead to a broad region of bunching at strong coupling as illustrated in the middle panel of Fig. \ref{fig:fig2}(d). At ultrastrong coupling, photon number conservation is reduced to conservation of photon parity due to the presence of counter-rotating terms. Hence, the cavity ground state [Eq. \ref{eq:gs}] has a finite probability for hosting two photons. The $|2,0\rangle$ component of the ground state can be stimulated to states with $|3,0\rangle$ component, which then radiates twice to end in the $|f\rangle \approx|1,0\rangle$, leaving the system in an excited state. Similarly, the $OIO$ and $OOI$ processes contribute to bunching at ultrastrong coupling, but to a lesser extent than $IOO$. Stimulated emission is a generic feature of ultrastrong coupling, but the bunching/antibunching near the gap is distinctly due to the Higgs polariton.

        The remaining processes, illustrated in the bottom panel of Fig. \ref{fig:fig2}(d) are negligible for $g^{(2)}$ even at ultrastrong coupling \footnote{See Supplemental Material Section \ref{sec:9x9} for a full decomposition of scattering pathways.}. A visible crescent in the bunching which lies above the anticrossing gap is enabled by the $OOII$ process and corresponds to a one-photon resonance between the mostly $|1,0\rangle$ intermediate state and the upper Higgs polariton state $|+\rangle$ which is mostly a mixture of $|2,0\rangle$ and $|0,1\rangle$. Specifically, as the ground state has a finite probability for hosting two photons, the scattering pathway between states with primary components $|2,0\rangle\overset{\hat{a}}{\to} |1,0\rangle \overset{\hat{a}}{\to} |0,0\rangle \overset{\hat{a}^\dag}{\to} |1,0\rangle \overset{\hat{a}^\dag}{\to} |+\rangle$ first emits these virtual photons towards the detector, then excites the polariton via injection of two input photons, resulting in a well-defined (if weak) resonance at the observed position. Other photon-parity allowed intermediate and final states contribute faintly.
        Finally, for very high-quality factor cavities, or at deep strong coupling other scattering processes can become relevant.

        In combination, the approach to ultrastrong coupling [Fig. \ref{fig:fig2}(a-c)] is characterized by a widening of the polaritonic gap (favoring antibunching from blocking two-photon transmission near $\omega_\mathrm{in} \approx \omega_\mathrm{cav} \approx \omega_h/2$) and increased stimulated emission of bunched dark-cavity photons. The minimum achievable $g^{(2)}$ (maximal antibunching) improves monotonically with light-matter coupling and cavity quality [Fig. \ref{fig:fig2}(e)].
        Crucially, measuring the statistics $g^{(2)}$ of transmitted photons is sensitive to the onset of ultrastrong coupling and dark-cavity photon fluctuations even when spectroscopic measurements of the polariton resonance only show weak changes from RWA predictions.

    \section{Applications to 2H-NbSe$_2$}
        While signatures of Higgs modes have been observed by terahertz spectroscopy in numerous superconductors \cite{tsuji2015theory,matsunaga2013higgs,matsunaga2014light,nakamura2020nonreciprocal,katsumi2020superconducting,chu2020phase,katsumi2020superconducting,katsumi2025distinct}, the Higgs typically lies close to $2\Delta$ and is obscured by pair-breaking excitations \cite{anderson1958random,devereaux2007inelastic,shimano2020higgs}. A promising material is 2H-NbSe$_2$, where superconductivity and charge-density-wave (CDW) order coexist and push the Higgs below $2\Delta$ \cite{littlewood1981gauge,littlewood1982amplitude,sooryakumar1980raman,measson2014amplitude,grasset2018higgs,du2025unveiling}.
        To model 2H-NbSe$_2$, we include a CDW mode $\hat{c}$
        \begin{align}
            \hat{H} &= \hbar\omega_\mathrm{cav} \hat{a}^\dag \hat{a} + \hbar\omega_h \hat{h}^\dagger \hat{h} + \hbar\omega_c \hat{c}^\dag \hat{c} + \kappa_{hc} (\hat{h}^\dagger \hat{c} + \hat{c}^\dagger \hat{h}) \notag\\ &\quad+ \kappa_h (\hat{a}+\hat{a}^\dag)^2(\hat{h}+\hat{h}^\dag) + \kappa_c (\hat{a}+\hat{a}^\dag)^2(\hat{c}+\hat{c}^\dag) 
        \end{align}
        and determine parameters for 2H-NbSe$_2$ by fitting experimental data to Ginzburg-Landau theory \footnote{See Supplemental Material Section \ref{sec:nbse2-params} for Ginzburg-Landau parameters.}. We phenomenologically include the quartic term $\kappa_4(\hat{a}+\hat{a}^\dag)^4$ to ensure the Hamiltonian remains bounded from below \footnote{See Supplemental Material Section \ref{sec:stability}.}.
        THz cavity quality factors range from $Q \sim 5$–$10$ in split-ring resonators and Fabry-Perot cavities \cite{appugliese2022breakdown,enkner2025tunable,mavrona2021thz,jarc2022tunable} to $Q \sim 10^3$ in photonic crystals and Tamm cavities \cite{yee2009high,zhang2016collective,xie2017terahertz,messelot2020tamm,tu2024tamm}; here we model with $Q\sim 120$.

        Fig.~\ref{fig:fig3} shows that the Higgs polariton characteristics---antibunching in the polariton gap, stimulated dark-cavity photon emission at ultrastrong coupling---persists in 2H-NbSe$_2$. A new feature arises from the second collective mode when the cavity mode is blue-detuned from the Higgs two-photon resonance, as hybridization of photon pairs and the CDW amplitudon generates a second two-photon polariton [Fig. \ref{fig:fig3}(b)].
        The input frequency for such higher-energy features, while below the linear-absorption pair breaking gap, lies above the two-photon resonance for pair-breaking excitations, which is expected to favor bunching and compete with the CDW-polariton-induced two-photon blockade. Still, the photon statistics from superconductor Higgs polaritons at lower input frequency remain robust in the experimentally relevant parameter regime, confirming that photon statistics are sensitive to ultrastrong coupling in realistic cavity quantum material. Avoided crossings in the two-photon manifold while the one-photon manifold is gapless will generically result in similar features.

        \begin{table}
            \centering
            \begin{tabular}{c|c|c|c|c|c}
                Type & Material & $T_c$ & $\hbar\omega_\mathrm{mode}/2$ & Irrep & Ref. \\\hline
                SC & 2H-NbSe$_2$ & 7.2 K & 1.2 meV & $E_{2g}$ & \cite{sooryakumar1980raman,measson2014amplitude} \\
                SC & 2H-TaS$_2$& 6.4 K & 0.62 meV & $E_{2g}$ & \cite{grasset2019pressure}\\
                \hline
                Bulk CDW & 1T-TaS$_2$ & 200 K & 5.0 meV & $A_{1g}$ & \cite{sugai1985lattice} \\
                & & & 7.1 meV & $A_{1g}$ & \cite{sugai1985lattice} \\
                Bulk CDW & 1T-TiSe$_2$ & 202 K & 4.8 meV & $E_g$ & \cite{uchida1981infrared} \\
                & & & 7.4 meV & $A_{1g}$ & \cite{uchida1981infrared} \\
                \hline
                1D CDW & o-TaS$_3$ & 220 K & 1.1 meV & $A_1$ & \cite{toda2009optical} \\
                1D CDW & (TaSe$_4$)$_2$I & 263 K & 5.6 meV & $A_{1}$ & \cite{zwick1985raman}\\
                1D CDW & (NbSe$_4$)$_{10}$I$_3$ & 289 K & 6.3 meV & $A_{1}$ & \cite{sekine1987raman} \\
                \hline 
            \end{tabular}
            \caption{Candidate materials for forming Higgs (amplitude) two-photon polaritons. Platforms include 2H TMDs with a superconducting phase, insulating 1T TMDs with a quasi-2D or 3D CDW with two amplitude modes, and quasi-1D materials which undergo a Peierls distortion. Mode frequencies are from Raman scattering experiments at low temperature; for 2H-TaS$_2$ values are at 6 GPa. $\hbar\omega_\mathrm{mode}/2$ is the frequency for a cavity to be two-photon resonant with the material mode. %1 meV = 0.2418 THz ($E=hf$) = 1.519 THz ($E=\hbar\omega$).
            }
            \label{tab:frequencies}
        \end{table}
    
    \section{Outlook}\label{sec:conclusion}
        We have shown that coupling THz cavity photon pairs to the Higgs mode in superconductors produces a two-photon nonlinearity in the ultrastrong coupling regime, whose signatures in the two photon coherence $g^{(2)}$---antibunching at the avoided two-photon resonance, stimulated emission of photons from the dark-cavity hybrid light-matter state---provide a diagnostic that is sensitive to the onset of of ultrastrong coupling.
        Besides superconductors, amplitude modes in cavity-embedded CDW compounds can similarly couple quadratically to THz cavity fields (see Table \ref{tab:frequencies} for candidates); such materials provide a rich playground to realize the two-photon Higgs polariton regime. This scenario is further enriched by multiple coupled collective modes \cite{ojeda2025fluctuation} or with nonlinear phononics \cite{mankowsky2016non,subedi2021light,juraschek21}.
        
        An intriguing future direction is the possibility to use such THz single-photon nonlinearities from cavity quantum materials as a route to extend quantum photonics from microwave or optics to the THz regime. For instance, integrating coherent matter excitations with high-quality THz cavities could be used to create squeezed and cat states \cite{flamini2019photonic}, or implement controlled multi-photon gates for THz sensing and information processing \cite{talkington2026input}. Finally in dissipative superconducting systems exotic band geometries and responses may be realized \cite{iemini2016dissipative,talkington2022dissipation,talkington2024linear,nava2023lindblad}, where the role of quantum geometry in general superconductors \cite{torma2022superconductivity,oh2025role} and the candidate material NbSe$_2$ in particular are of recent interest \cite{yu2025quantum}.\\

    \section{Acknowledgments}
        S.T. acknowledges support from the NSF under Grant No. DGE-1845298. M.C. and B.K. acknowledge support from Charles E. Kaufman Foundation under a New Initiative grant, the Alfred P. Sloan Foundation through a Sloan Research Fellowship, and the Center for Quantum Information, Engineering, Science and Technology (QUIEST) of the University of Pennsylvania. We thank M. Hafezi, K. Katsumi, A. Srivastava, B.F. Mead, G. Nambiar, and the Penn CMT Group for discussions surrounding this work.

    %\bibliography{references}

%%%%%% SUPPLEMENTARY MATERIAL %%%%%
    
    \supplement
    % \abstract{...}
        
        \supplementtoc
        
    \section{Non-Markovian Input-Output Relations}\label{sec:io-theory}

        In this section, we describe non-Markovian input-output relations for THz cavity quantum materials at ultrastrong coupling. These relations connect the detectable output fields to cavity photons and input fields, and must in general be solved together with a Langevin equation for the cavity photons which depends on the input field as well. The latter is straightforward for free Hamiltonians \cite{ciuti2005quantum}, however poses a significant challenge for a non-linear many-body Hamiltonian of the cavity and quantum material. In section \ref{sec:scattering}, we describe a scattering matrix formalism which permits the straightforward computation of output photon correlation functions in terms of multi-point cavity photon correlation functions that can be computed numerically.
        
        Consider a single-mode cavity coupled to a ``left'' ($L$) and ``right'' ($R$; location of the detector) photon bath, described by a Hamiltonian (with $\hbar=1$ from here on):
        \begin{align}
            \hat{H} ~&=~ \hat{H}_0 + \hat{V}  \label{eq:Hsysbath} \\
            \hat{H}_0 ~&=~ \hat{H}_{\rm cav} ~+~ \int_0^\infty d\omega~ \omega \left( \hat{b}^\dag_{L}(\omega) \hat{b}_{L}(\omega) + \hat{b}^\dag_{R}(\omega) \hat{b}_{R}(\omega) \right) \\
            \hat{V} ~&=~ \sum_{\alpha = L,R} \int_0^\infty d\omega~ \sqrt{\frac{\gamma_{\omega}}{2\pi}} \left( \hat{a}^\dag \hat{b}_{\alpha}(\omega) + \hat{a} \hat{b}^\dag_{\alpha}(\omega)\right)
        \end{align}
        Here, $\hat{H}_{\rm cav}$ describes an arbitrary single-mode cavity with a  photon mode $\hat{a}$ that is coupled to an interacting quantum material. The fields $\hat{b}_L(\omega)$ and $\hat{b}_R(\omega)$ describe two continuous photon baths $L$ and $R$. Finally, $\hat{V}$ describes the tunneling of photons between the cavity and the bath, which is parameterized via a frequency-dependent tunnel rate $\gamma_\omega$. Crucially, the photon bath only has modes with positive frequencies $\omega > 0$.
        
        One can now derive an input-output relation for the bath photons that relates an initial state at time $t_i \to -\infty$ to a final state at time $t_f \to +\infty$ \cite{gardiner1985input}. Consider the Heisenberg equation of motion for the bath photons
        \begin{align}
            \partial_t \hat{b}_{\alpha}(\omega; t) &= -i \omega \hat{b}_{\alpha}(\omega; t) - i \sqrt{\frac{\gamma_{\omega}}{2\pi}} \hat{a}(t)  \label{eq:inputOutputFrequencyDomain}
        \end{align}
        where $\hat{b}_{\alpha}(\omega; t)$ represents the Heisenberg-picture bath photon at time $t$ with frequency $\omega$. Integrating the equation of motion, one obtains
        \begin{align}
            \hat{b}_{\mathrm{out},\alpha}(\omega) = \hat{b}_{\mathrm{in},\alpha}(\omega) - i \sqrt{\frac{\gamma_\omega}{2\pi}} \int\limits_{t_i}^{t_f} \frac{dt}{2\pi} ~e^{i\omega t} \hat{a}(t)
        \end{align}
        with input ($\hat{b}_{\mathrm{in},L}$ and $\hat{b}_{\mathrm{in},R}$; distant past) and output ($\hat{b}_{\mathrm{out},L}$ and $\hat{b}_{\mathrm{out},R}$; distant future) scattering fields are defined as \cite{gardiner1985input}
        \begin{align}
            \hat{b}_{\mathrm{in},\alpha}(\omega) &\equiv \hat{b}_{\alpha}(\omega; t = t_i)~ e^{i \omega t_i}\\
            \hat{b}_{\mathrm{out},\alpha}(\omega) &\equiv \hat{b}_{\alpha}(\omega; t = t_f)~ e^{i \omega t_f}
        \end{align}
        In principle, the Fourier-transformed real-time input and output fields
        \begin{align}
            \hat{b}_{\mathrm{in},\alpha}(t) &= \int_0^\infty d\omega~ e^{-i\omega t} \hat{b}_{\mathrm{in},\alpha}(\omega) \\\hat{b}_{\mathrm{out},\alpha}(t) &= \int_0^\infty d\omega~ e^{-i\omega t} \hat{b}_{\mathrm{out},\alpha}(\omega)  \label{eq:bout}
        \end{align}
        now permit computing photon counts and correlations at the detector. These real-time fields satisfy a non-Markovian input-output relation 
        \begin{align}
            \hat{b}_{\mathrm{out},\alpha}(t) = \hat{b}_{\mathrm{in},\alpha}(t) - i \int \frac{dt'}{2\pi}\ \Gamma(t-t') \hat{a}(t') 
        \end{align}
        where
        \begin{align}
            \Gamma(t-t') = \int_0^\infty d\omega\ e^{-i\omega(t-t')} \sqrt{\gamma_\omega}
        \end{align}
        and, importantly, care must be taken to solely integrate over positive bath frequencies.
        
        In conventional AMO settings, a Markov approximation is often employed where the system-bath coupling $\gamma_\omega \to \gamma$ is taken to be frequency-independent, and importantly, the photon bath is assumed to extend across negative frequencies \cite{gardiner1985input}. In this case, $\Gamma(t-t') \to \sqrt{\gamma} \delta(t-t')$ yields a simple time-local relation between input, output, and cavity fields $\hat{b}_{\mathrm{out},R}(t) = \hat{b}_{\mathrm{in},R}(t) - i \sqrt{\gamma} \hat{a}$, which must be solved in conjunction with a Langevin equation for the cavity photons $\partial_t \hat{a}(t) = i [ \hat{H}_\mathrm{cav}, \hat{a}(t) ] - \gamma \hat{a}(t) - i\sqrt{\gamma} [\hat{b}_{\mathrm{in},L}(t) + \hat{b}_{\mathrm{in},R}(t)]$ \cite{gardiner1985input}. 
        
        However, the Markov approximation fails rather drastically at ultrastrong coupling \cite{ciuti2006input}: In dark cavities with a ground state with finite photon fluctuations, the Markovian input-output relation would predict a finite output photon number $\langle \hat{b}_\mathrm{out}^\dag(t) \hat{b}_\mathrm{out}(t) \rangle \sim \langle \hat{a}^\dag(t) \hat{a}(t) \rangle$, erroneously predicting that a dark cavity radiates without an input. To remedy this, it is essential to retain only positive-frequency bath modes and work with a non-Markovian input-output relation. In section \ref{sec:scattering}, we show that the output field correlation functions can be computed efficiently using a non-Markovian scattering matrix approach, which rewrites the output response as a sum of multi-time photon correlation function for the intra-cavity Hamiltonian, for classical THz input fields.
        
    \section{Weak Coupling Analytic Solution}\label{sec:analytic}

        The analytic solution of the driven-dissipative cavity RWA Lindbladian in the main text starts from the $P$-representation solution of Ref. \cite{drummond1980non}; in this work, Drummond, McNeil and Walls consider the master equation
        \begin{align}
            \hbar\partial_t \hat\rho = -i[\hat H(t),\hat \rho] + \gamma \mathcal{D}[\hat a,\hat \rho] + \gamma_h \mathcal{D}[\hat h,\hat \rho]/2
        \end{align}
        with dissipators $\sqrt{\gamma} \hat a$ and $\sqrt{\gamma_h/2} \hat h$ and a driven Hamiltonian
        \begin{align}
        \hat H &= \hbar\omega_\mathrm{cav} \hat a^\dag \hat a + \hbar\omega_h \hat h^\dag \hat h + i \kappa(\hat a^\dag \hat a^\dag \hat h - \hat a\hat a\hat h^\dag)\notag\\ &+ i(f_\mathrm{cav} \hat a^\dag e^{-i\omega_\mathrm{in}t} - \bar{f}_\mathrm{cav} \hat ae^{i\omega_\mathrm{in}t})\notag\\ &+ i(f_h \hat h^\dag e^{-2i\omega_\mathrm{in}t} - \bar{f}_h \hat h e^{2i\omega_\mathrm{in}t})
        \end{align}
        This is precisely the RWA master equation of the main text if we consider a gauge transformation ($\hat h^\dag\to i\hat h^\dag$, $\hat a^\dag\to -i\hat a^\dag$) and $f_h\to 0$. Ref. \cite{drummond1980non} uses the generalized $P$-representation to write an equation $\partial_t P(\alpha_\mathrm{cav},\alpha_h)=O(\alpha_\mathrm{cav},\alpha_h)P(\alpha_\mathrm{cav},\alpha_h)$ for a specific differential operator $O$, then takes the limit that $\gamma_h\gg 2\gamma$ to adiabatically eliminate $\alpha_h$ by replacing it with its effective value $(f_h-\kappa\alpha_\mathrm{cav}^2)/(\gamma_h/2+i\delta_h)$. The equation is then a Fokker-Planck equation of one complex variable that is directly solvable by the method of potentials which yields an expression for $P$.
        
        Remarkably the expressions for the normally-ordered correlation functions
        \begin{align}
        \langle (a^\dag)^n a^{n'}\rangle = \int d\alpha_\mathrm{cav} d\bar{\alpha}_\mathrm{cav}\ \bar{\alpha}^n \alpha^{n'} P(\alpha)
        \end{align}
        are all integrable and yield
        \begin{align}
            I_{nn'} &= \sum_{m=0}^\infty\bigg[\frac{2^m}{m!} \left(-\sqrt{f_h/\kappa}\right)^{m+n} \left(-\sqrt{\bar{f}_h/\kappa}\right)^{m+n'} \notag\\ &\quad {_2}F_1(-(m+n),j_1,j_2,2)\ {_2}F_1(-(m+n'),\bar{j}_1,\bar{j}_2,2)\bigg]
        \end{align}
        for Gaussian hypergeometric functions ${_2}F_1$, and $j_1 = (\gamma+i\delta_\mathrm{cav})(\gamma_h/2+i\delta_h)/2\kappa^2 + f_\mathrm{cav}(\gamma_h/2+i\delta_h)/(2\kappa\sqrt{\kappa f_h})$, $j_2 = (\gamma+i\delta_\mathrm{cav})(\gamma_h/2+i\delta_h)/\kappa^2$, and detunings $\delta_\mathrm{cav}=\hbar\omega_\mathrm{cav}-\hbar\omega_\mathrm{in}$, $\delta_h=\hbar\omega_h-2\hbar\omega_\mathrm{in}$. If $f_\mathrm{cav}$ is too big the series will diverge with increasing $m$ while if $f_\mathrm{cav}$ is small the first term gives a very good approximation to $I_{nn'}$. Now if we simplify to $f_h$ real and positive (or approaching zero), and specify $n'=n$ then this simplifies to
        \begin{align}
            I_{nn} = \sum_{m=0}^\infty\bigg[\frac{2^m}{m!} 
            \left(\frac{f_h}{\kappa}\right)^{m+n}\ &{_2}F_1(-(m+n),j_1,j_2,2)\notag\\& {_2}F_1(-(m+n),\bar{j}_1,\bar{j}_2,2)\bigg]
        \end{align}
        Taking the $f_h\to 0$ limit we see that there is a $f_h^{m+n}$ term, and $j_1\sim 1/\sqrt{f_h}$, but the dominant contribution to $_2F_1$ is $j_1^{m+n}$, or $f_h^{-(m+n)/2}$; with two $_2F_1$ terms, these cancel to yield a finite limit. Specifically
        \begin{align}
        I_{nn} = |\phi|^{2n} \bigg|\frac{\Gamma(j)}{\Gamma(j+n)}\bigg|^2 {_0}F_2(j+n,\bar{j}+n;2|\phi|^2)
        \end{align}
        where ${_0}F_2$ is a generalized confluent hypergeometric function, $j = j_2$, and $\phi=-if_\mathrm{cav}(\gamma_h/2+i\delta_h)/\kappa^2$.

\begin{figure}
            \centering
            \includegraphics[width=0.8\linewidth]{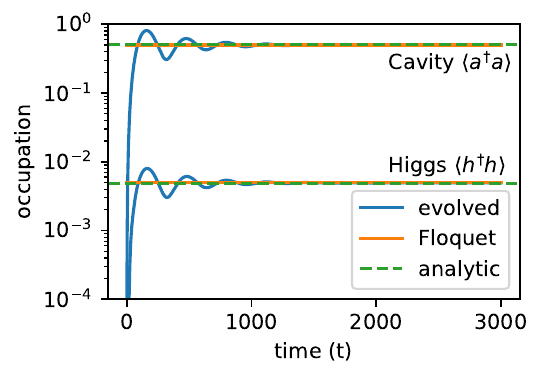}
            \caption{Comparison of methods for finding the steady-state density: time-evolution, Floquet null-space determination, and $P$-representation analytics. Top line is the occupation of the cavity mode, while the bottom line is the occupation of the Higgs mode. The methods agree as they should, giving us confidence in our results.}
            \label{fig:s1}
        \end{figure}

        Fig. \ref{fig:s1} depicts the analytic solution vs the exact numerical dissipative steady state (computed via evolving the Lindblad equation using an efficient method detailed in Ref. \cite{talkington2020efficient} or via Floquet solution of the steady state \cite{chen2024periodically}), for $\omega_\mathrm{cav}=0.5$, $\omega_h=1$, $\omega_\mathrm{in}=0.5$, $\kappa=0.1$, $\gamma=0.001$, $\gamma_h=0.2$, and $f=0.01$, and using a converged bosonic mode cutoff in photons and Higgs excitations. The exact steady state density correlation function is well reproduced via the analytic expression at strong coupling in the RWA. The ultrastrong coupling regime of the main text instead breaks the RWA and requires combining a numerical solution with a non-Markovian scattering matrix approach for output fields, as described in the main text.

        \label{section:usc-to-sc}
        \subsection{Comparing RWA and exact solutions for the second-order photon coherence at weak coupling}

\begin{figure}
                \centering
                \includegraphics[width=0.7\linewidth]{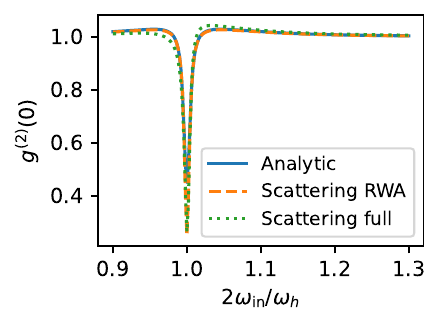}
                \caption{Comparison of $g^{(2)}(0)$ as a function of $\omega_\mathrm{in}$ at $\kappa_h=0.01$, $\omega_\mathrm{cav}=0.5$, $\omega_h=1$, $\gamma=0.001$, $\gamma_h=0.2$, and $f=0.001$. For the scattering-matrix approach we take $N_\mathrm{cav}=N_h=6$. Note the strong agreement between scattering RWA approach and the analytic result.}
                \label{fig:s2}
            \end{figure}

            Fig. \ref{fig:s2} compares $g^{(2)}(0)$ from three calculations: (1) an exact non-Markovian input-output scattering matrix calculation for the full Higgs-polariton Hamiltonian with counter-rotating terms [Eq. \ref{eq:H}] (which correctly treats the positivity of bath photon mode frequencies); (2) using the analytic RWA solution with the Markovian input-output relations of Gardiner and Collett \cite{gardiner1985input}, and (3) a non-Markovian input-output calculation starting from the RWA Hamiltonian. We observe that the RWA results agree precisely, which is rooted in a trivial RWA dark-cavity state with a vacuum ground state. In contrast, even for modest light-matter coupling strengths $\kappa_h / \hbar\omega_h = 0.01$ that are far from ``ultrastrong'' coupling,  the exact computation of $g^{(2)}$ exhibits deviations from the RWA Hamiltonian as a result of its additional counter-rotating terms. In Section. \ref{sec:lindblad-limit} we analytically derive the limit under which the non-Markovian scattering matrix theory reduces to the Markovian RWA result.

        \section{Temperature Dependent Ginzburg-Landau Parameters} \label{sec:TDGL}
    
        Consider $T$-independent $\beta$, $m^*$, $q$, and
        $\alpha(T) = \alpha_0(1-T/T_c)$.
        Gorkov tells us \cite{gorkov1959microscopic} that when we choose a normalization so that $\Delta$ is the superconducting gap that $\alpha_0=N(0)$ and
        \begin{align}
        \beta = \frac{7\zeta(3) N(0)}{16\pi^2(k_BT_c)^2}\label{eq:beta-gorkov}
        \end{align}
        where $N(0)$ is the single-spin density of states at $\epsilon_F$. The gap is then
        \begin{align}
        2\Delta(T) &= 2\sqrt{\alpha(T)/2\beta}\\
        &= 2\pi \sqrt{8/7\zeta(3)} k_B T_c \sqrt{1-T/T_c}\label{eq:GL-gap}
        \end{align}
        where $\zeta(3)\approx1.20206$ is Apery's constant.
        Gorkov also tells us $K=(7\zeta(3)/48\pi^2) N(0) (\hbar v_F/k_B T_c)^2$.
        The superfluid density is $\rho_s = 2m^* K|\Delta|^2/\hbar^2$, or $\rho_s=(1/3)N(0) m^* v_F^2 (1-T/T_c)$.

        Now, the standard $T$-dependent weak coupling BCS gap is
        \begin{align}
        2\Delta(T) &= 2\pi e^{-\gamma_E} k_B T_c \, \tanh\big(e^{\gamma_E} \sqrt{8/7\zeta(3)} \sqrt{T_c/T-1}\big)
        \end{align}
        where $\gamma_E\approx0.577216$ is the Euler-Mascheroni constant.
        Near $T_c$ this can be approximated by
        \begin{align}
        2\Delta(T)
        &\approx 2\pi \sqrt{8/7\zeta(3)}\, k_B T_c \sqrt{1-T/T_c}\label{eq:BCS-gap}
        \end{align}
        where we took the first term of the Taylor series of tanh and approximated $T$ to $T_c$ in the denominator.
        We see that with the Gorkov choice of $\alpha$ and $\beta$ the weak-coupling BCS theory and the GL theory agree (the gap is the same in Eq. \ref{eq:GL-gap} and Eq. \ref{eq:BCS-gap}).
        
        Now what happens at $T$ substantially less than $T_c$ over the full temperature range?
        We let $\beta_\mathrm{full}$ assume its Gorkov value at $T_c$, Eq. \ref{eq:beta-gorkov}, and
        \begin{align}
        \alpha_\mathrm{full}(T) = (7\zeta(3) e^{-2\gamma_E}/8) N(0) \tanh^2(e^{\gamma_E}\sqrt{8/7\zeta(3)} \sqrt{T_c/T-1})
        \end{align}
        so that
        $
        2\Delta(T) = 2 \sqrt{\alpha_\mathrm{full}(T)/2\beta_\mathrm{full}}
        $
        which recovers the weak-coupling BCS gap and provides us with temperature dependent Ginzburg-Landau parameters.

        Working in SI units, Eq. (\ref{eq:free}) is a free energy density with units of $\mathrm{J/m^3}$. Consequently, $[\Delta]=\mathrm{J}$, $[\alpha]=1/\mathrm{(J\cdot m^3)}$, $[\beta]=1/\mathrm{(J^3\cdot m^3)}$.
        The gradient term $[K]=1/\mathrm{J\cdot m}$.
        Finally the inertial coupling term $P$ has units $1/\mathrm{(J^3\cdot m^3)}$.

    \section{Physical Parameters for 2H-NbSe$_2$}\label{sec:nbse2-params}

        Consider a two-level approximation of coupling a Higgs mode at $\hbar\omega_h=2\Delta$ to a charge density wave amplitudon at $\hbar\omega_c$
        \begin{align}
            W = \begin{pmatrix}
            \hbar\omega_c(T) & \kappa_{hc}(T)\\
            \kappa_{hc}(T) & 2\Delta(T)
            \end{pmatrix}
        \end{align}
        In 2H-NbSe$_2$, the CDW transition temperature is near 33 K, so for $T\leq 7.2~\mathrm{K}=T_c$, $\omega_c$ is $T$-independent. We assume that $2\Delta(T)$ assumes the value $4.062\, k_B T_c \, t(T)$ where $t(T) \equiv \tanh\big(1.737 \sqrt{T_c/T-1}\big)$. This leading coefficient is slightly larger than the weak-coupling BCS coefficient of 3.528 but the temperature scaling still has the same form \cite{clayman1971superconducting}. To find the $T$-dependence of $\kappa_{hc}$, we note that the lowest order term in a two-component GL free energy for CDW and superconductivity that can couple the two fields is $\kappa_{hc}^0 |\Delta_h|^2 |\Delta_c|^2$. Expanding about both minima yields $\kappa_{hc}^0 |\Delta_h|_0 |\Delta_c|_0 \Delta_h' \Delta_c'$. Here, $|\Delta_h|_0$ scales as $t(T)$, and $|\Delta_c|_0$ is approximately temperature independent in the superconducting phase. Assuming that $\kappa_{hc}^0$ is temperature independent, $\kappa_{hc}$ scales as $t(T)$. We now have
        \begin{align}
            W = \begin{pmatrix}
            \hbar\omega_c & \kappa_{hc}^0 t(T)\\
            \kappa_{hc}^0 t(T) & 2\Delta(T)
            \end{pmatrix}
        \end{align}
        which has eigenvalues
        \begin{align}
            \hbar\omega_\pm(T) &= \frac{\hbar\omega_c + 2\Delta(T)}{2}\notag\\&\qquad \pm \frac{1}{2}\sqrt{[2\kappa_{hc}^0t(T)]^2 + (\hbar\omega_c-2\Delta(T))^2}
        \end{align}
        Raman measurements observe the dressed resonances $\hbar\omega_\pm$ rather than $\hbar\omega_c$ and $\hbar\omega_h$. By extracting $\hbar\omega_\pm(T^*)$ and the pair-breaking gap at temperature $T^*$, the bare Higgs and CDW amplitudon frequencies and the Higgs-CDW coupling can be extracted by solving
        \begin{align}
            \hbar\omega_c &= \hbar\omega_+(T^*)+\hbar\omega_-(T^*) - 2\Delta(T^*) \\
            \kappa_{hc}^0 &= \frac{\sqrt{[\hbar\omega_+(T^*)\!-\!\hbar\omega_-(T^*)]^2 \!-\! (\hbar\omega_c\!-\!2\Delta(T^*))^2}}{2t(T^*)}
        \end{align}
        Parameters at other temperatures can then be extrapolated in principle via the known temperature scalings. We ignore the CDW phason which can couple in linear response \cite{gruner1988dynamics}, as it lies at much lower energy in 2H-NbSe$_2$ \cite{sheng2024terahertz}. Other sub-gap modes such as the Bardasis-Schrieffer mode, while of recent interest, are not expected here \cite{matsumoto2025new}.

        \begin{figure}
            \centering
            \includegraphics[width=0.5\linewidth]{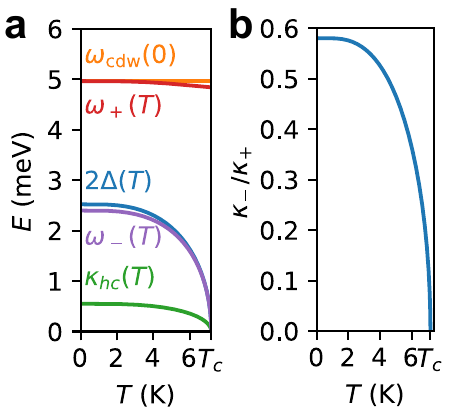}
            \caption{Temperature dependent \textbf{(a)} collective mode and \textbf{(b)} coupling parameters for 2H-NbSe$_2$.}
            \label{fig:s3}
        \end{figure}

        In Ref. \cite{measson2014amplitude}, Raman measurements at 2K find 19.2/cm and 40.0/cm (2.38 meV and 4.96 meV) for $\omega_\pm$ so that $\hbar\omega_c=4.84$ meV, $\kappa_{hc}^0=0.546$ meV. Here we note that $2\Delta(T)=4.062\, k_B T_c t(T)$ so that $2\Delta(0)=2.52$ meV in agreement with experiments on the superconducting gap \cite{huang2007experimental}.

        The relative scale of the non-linear light-matter coupling to Higgs vs CDW amplitudon modes can be estimated by comparing to terahertz third harmonic generation \cite{feng2023dynamical}. The ratio can be written as
        \begin{align}
            \frac{\kappa_-(T)}{\kappa_+(T)} = \frac{\kappa_c(T)|\langle-|c\rangle|^2 + \kappa_h(T)|\langle-|h\rangle|^2}{\kappa_c(T)|\langle+|c\rangle|^2 + \kappa_h(T)|\langle+|h\rangle|^2}
        \end{align}
        where $|\!\pm\!(T)\rangle$ are the eigenvectors corresponding to $\hbar\omega_\pm(T)$. Assuming that $\kappa_c(T)=\kappa_c^0$, $\kappa_h(T)=\kappa_h^0 t(T)$ we see that by rearranging we can solve for the ratio $\kappa_h^0/\kappa_c^0$ using the known expressions for $t(T)$, $|\pm(T)\rangle$, $|h\rangle=(0,1)$, $|c\rangle=(1,0)$, and $\kappa_-(T^*)/\kappa_+(T^*)$ at a known temperature $T^*$. Using the data from Ref. \cite{feng2023dynamical} at 4.5 K we find $\kappa_h^0/\kappa_c^0=0.5803$. This then enables us to find $\kappa_h$ and $\kappa_c$ at all temperatures, with just the overall coupling scale undetermined. Fig. \ref{fig:s3} depicts the resulting estimates for temperature-dependent model parameters for NbSe\textsubscript{2}.

        \begin{figure}[!b]
            \centering
            \includegraphics[width=0.7\linewidth]{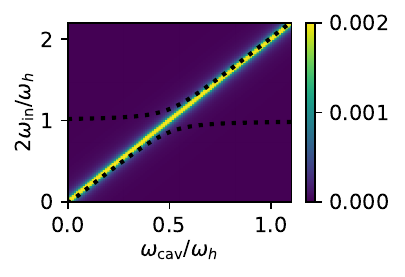}
            \caption{Strong coupling transmission in the RWA regime, for parameters corresponding to Fig. \ref{fig:fig1} of the main text.}
            \label{fig:s4}
        \end{figure}

        \begin{figure*}
            \centering
            \includegraphics[height=1.75 in]{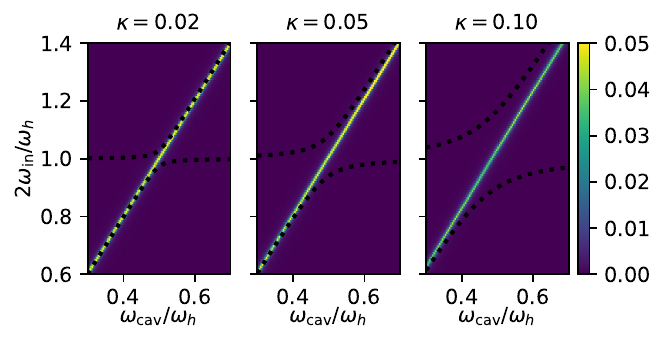}\quad
            \includegraphics[height=1.75 in]{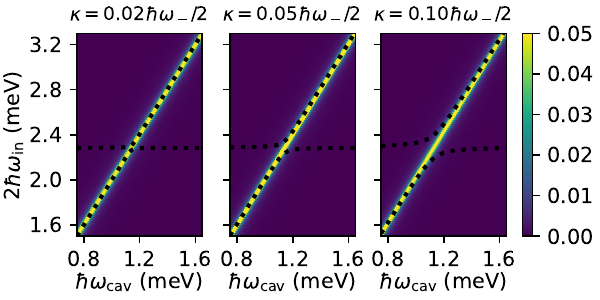}
            \caption{\textbf{(left)} Ultrastrong coupling transmission beyond the RWA, corresponding to Fig. \ref{fig:fig2} of the main text. \textbf{(right)} Ultrastrong coupling transmission beyond the RWA, corresponding to Fig. \ref{fig:fig3} of the main text.}
            \label{fig:s5}
        \end{figure*}

    \section{Stability Criterion}
        \label{sec:stability}

        The large photon-displacement behavior of Hamiltonians with cubic nonlinearities can be regularized via a small positive quartic term $\sim (a + a^\dag)$ \cite{ojeda2024equilibrium}. We choose the prescription where $\kappa_4$ is the smallest value that ensures stability.

        Consider first the case of one collective mode 
        \begin{align}
            \hat{H} &= \hbar \omega_\mathrm{cav} \hat{a}^\dagger \hat{a} + \kappa_4 (\hat{a} + \hat{a}^\dag)^4 \notag\\ &\quad+ \hbar \omega_h \hat{h}^\dagger \hat{h} + \kappa(\hat{a}+\hat{a}^\dag)^2(\hat{h}+\hat{h}^\dag) 
        \end{align}
        Since $h$ and $a$ commute, the Hamiltonian can be written as a sum of positive semidefinite terms:
        \begin{align}
        \hat{H} &= \hbar \omega_\mathrm{cav} \hat{a}^\dagger \hat{a} + (\kappa_4-\frac{\kappa^2}{\hbar\omega_h}) (\hat{a} + \hat{a}^\dag)^4 \\ &\quad+ \hbar\omega_h (\hat{h}^\dag + \frac{\kappa}{\hbar\omega_h}(\hat{a}+\hat{a}^\dag)^2)(\hat{h} + \frac{\kappa}{\hbar\omega_h}(\hat{a}+\hat{a}^\dag)^2)  \notag
        \end{align}
        One can read off that stability is guaranteed provided $\kappa_4\geq \kappa^2/\hbar\omega_h$.

\begin{figure}
            \centering
            \includegraphics[width=0.8\linewidth]{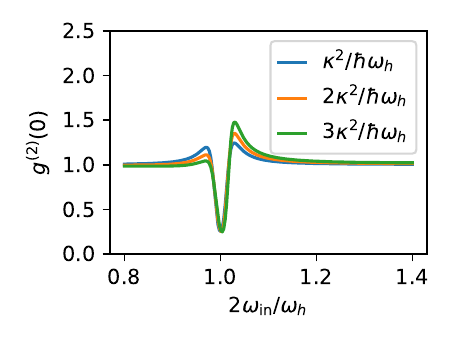}
            \caption{Varying the phenomenological stabilization parameter $\kappa_4$ for the same parameters as Fig. \ref{fig:fig2}(a) at $\omega_\mathrm{cav}/\omega_h=0.5$. The peak heights change slightly with $\kappa_4$, but the polariton splitting structure remains unchanged.}
            \label{fig:vary-kappa4}
        \end{figure}

        Now, in Fig. \ref{fig:vary-kappa4} we see that the effect of varying the phenomenological value of $\kappa_4$ is primarily quantitative and the qualitative features of the polariton splitting remain even as $\kappa_4$ increases.

        Consider now the Higgs-CDW Hamiltonian of the main text:
        \begin{align}
            \hat{H} &= \hbar \omega_\mathrm{cav} \hat{a}^\dagger \hat{a} + \hbar \omega_h \hat{h}^\dagger \hat{h} + \hbar \omega_c \hat{c}^\dag \hat{c} + \kappa_4 (\hat{a} + \hat{a}^\dag)^4 \notag\\&\quad + \kappa_h(\hat{a}+\hat{a}^\dag)^2(\hat{h}+\hat{h}^\dag) + \kappa_c(\hat{a}+\hat{a}^\dag)^2(\hat{c}+\hat{c}^\dag) 
        \end{align}
        For this case it is useful to recall the more general relation (where $\bm{\kappa}$'s components are Hermitian)
        \begin{align}
        \bm{x}^\dag A \bm{x} + \bm{\kappa}^\top \bm{x} + \bm{x}^\dag \bm{\kappa} &= (\bm{x}^\dag + \bm{\kappa}^\top A^{-1})A(\bm{x}+A^{-1}\bm{\kappa})\notag\\&\quad - \bm{\kappa}^\top A^{-1}\bm{\kappa}
        \end{align}
        where for our case $\bm{x}=(\hat{h},\hat{c})$ with $\bm{\kappa}=(\hat{a}+\hat{a}^\dag)^2\cdot(\kappa_h,\kappa_c)$ and $A=\begin{pmatrix}\hbar\omega_h & \kappa_{hc}\\\kappa_{hc}&\hbar\omega_c\end{pmatrix}$.
        Inserting the last term into $\hat{H}$ we see that the quartic term is
        \begin{align}
        &\kappa_4 (\hat{a} + \hat{a}^\dag)^4 - \bm{\kappa}^\top A^{-1}\bm{\kappa}
        = (\hat{a} + \hat{a}^\dag)^4 \bigg[\kappa_4- \frac{1}{\hbar^2\omega_h\omega_c - \kappa_{hc}^2} \cdot \notag\\ &\hspace{1 in} \begin{pmatrix}\kappa_h\\\kappa_c\end{pmatrix}^\top\begin{pmatrix}\hbar\omega_c & -\kappa_{hc}\\-\kappa_{hc}&\hbar\omega_h\end{pmatrix}\begin{pmatrix}\kappa_h\\\kappa_c\end{pmatrix}\bigg] 
        \end{align}
        where we took the matrix inverse explicitly, and so contracting the matrix with the vectors, the stability criterion is
        \begin{align}
        \kappa_4 \geq \frac{\hbar\omega_c\kappa_h^2 - 2 \kappa_h \kappa_c \kappa_{hc} + \hbar\omega_h\kappa_c^2}{\hbar^2\omega_h\omega_c - \kappa_{hc}^2}
        \end{align}
        There is an additional stability criterion that $A\geq 0$ so that the $h/c$ subspace itself is stable. This is satisfied by the parameters we consider and must be satisfied for any physical model. In the main text we take $\kappa_4$ to saturate this inequality.

    \section{Ground State Occupations}\label{sec:GS-fock}

\begin{figure*}
            \centering
            \includegraphics[width=\linewidth]{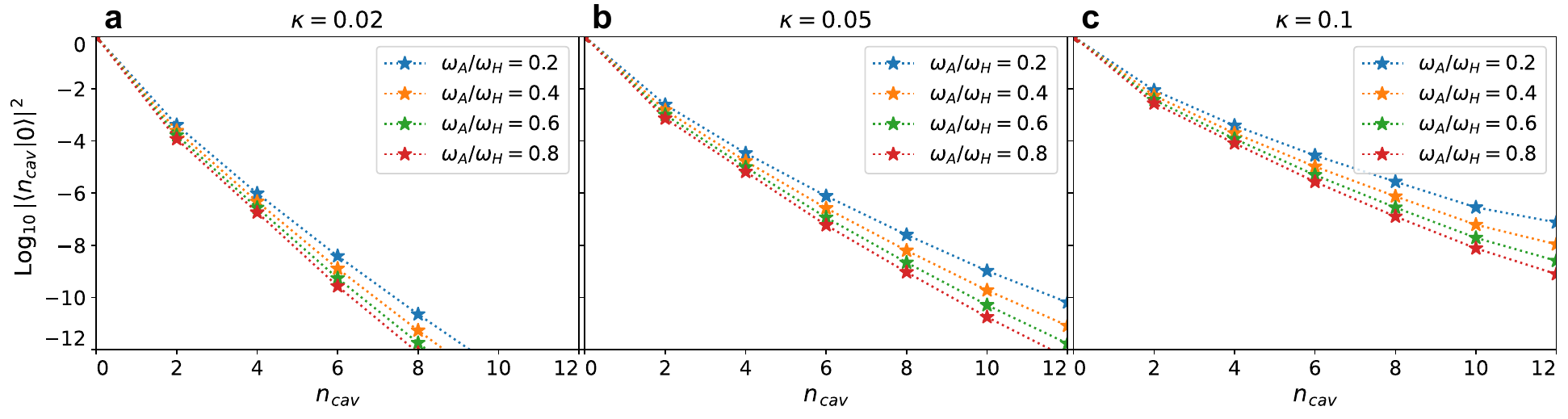}
            \caption{Ground state photon number overlaps for the same parameters as Fig. \ref{fig:fig2}. The occupations fall off rapidly with $n_{cav}$, but are non-zero and are enhanced when $\kappa$ becomes large at ultrastrong coupling. The occupancies of odd number states are precisely zero by the photon parity symmetry of the Hamiltonian.}
            \label{fig:GS-fock}
        \end{figure*}

        As discussed in the main text, at ultrastrong light-matter coupling, the dark-cavity ground state $|0\rangle$ has finite support on 2, 4, 6, ... photon sectors. Here we compute this numerically and plot in Fig. \ref{fig:GS-fock} for the same parameters as Fig. \ref{fig:fig2}. We see that there is a rapid fall-off with $n_{cav}$, but that the $n$-photon occupancies are non-zero and grow as $\kappa$ becomes larger.

    \section{Transmission / $G^{(1)}$}\label{sec:transmission}

        Here we will consider the transmission of photons
        \begin{align}
        T = \frac{\sum_{\omega_\mathrm{out}} G^{(1)}_{\mathrm{out},\omega_\mathrm{out}}}{\sum_{\omega_\mathrm{in}} G^{(1)}_{\mathrm{in},\omega_\mathrm{in}}}
        \end{align}
        rather than the transmission of power since it is the more natural quantity to consider when photons are counted independent of frequency (as is the case in superconducting single photon detectors). Now, the denominator of this quantity is $\sum_{\omega_\mathrm{in}} G^{(1)}_{\mathrm{in},\omega}$ which for either a monochromatic coherent input (as in weak coupling/RWA analysis), or for a one/two-photon Fock state input (as in the scattering matrix analysis), the result is a constant independent of input frequency and cavity parameters. This means that for the setups we consider
        \begin{align}
        T/s = \sum_{\omega_\mathrm{out}} G^{(1)}_{\mathrm{out},\omega_\mathrm{out}}
        \end{align}
        with an overall scale $s$ set by the normalization of the input state. Let us now plot this rescaled transmission for each of the cases in the main text as photon counting transmitted photons could be a useful experimental measure too. Doing so in Fig. \ref{fig:s4} and \ref{fig:s5} we observe that the transmission is essentially the resonance condition $\omega_\mathrm{in}=\omega_\mathrm{cav}$.

    \begin{widetext}

    \section{Scattering Matrix Input-Output Theory at Ultrastrong Coupling}\label{sec:scattering}

        \subsection{Scattering Matrix Approach}

            We now formulate an exact scattering matrix approach for computing the one-photon and two-photon right-side output correlation functions at the detector
            \begin{align}
                G^{(1)}(t_m) &= \langle \hat{b}_{\mathrm{out},R}^\dag(t_m) \hat{b}_{\mathrm{out},R}(t_m)\rangle \\
                G^{(2)}(t_m, t) &= \langle \hat{b}_{\mathrm{out},R}^\dag(t_m) \hat{b}_{\mathrm{out},R}^\dag(t_m + t) \hat{b}_{\mathrm{out},R}(t_m + t) \hat{b}_{\mathrm{out},R}(t_m)\rangle
            \end{align}
            in transmission geometry, by perturbatively expanding Eq. (\ref{eq:Hsysbath}) in photon tunneling processes from the ``left'' (input) to the ``right'' (detector).
            Let us first consider single-photon detection
            \begin{align}
            G^{(1)}(t_m) &= \langle \psi_i(t_f)|\hat b_{\mathrm{out},R}^\dag(t_m) \hat b_{\mathrm{out},R}(t_m)|\psi_i(t_f)\rangle
            \end{align}
            where the scattering state $|\psi_i(t_f)\rangle$
            at a final time $t_f$ must be computed from the asymptotic incoming
            scattering state $|\phi_i\rangle$ at time $t_i$. We insert a complete set of final states $F$
            \begin{align}
            G^{(1)}(t_m) = \sum_{F} |\langle F| \hat b_{\mathrm{out},R}(t_m)|\psi_i(t_f)\rangle|^2
            \end{align}
            We choose an incoming state in the distant past
            \begin{align}
            |\phi_i\rangle = |n_{\omega_\mathrm{in}}=n_\mathrm{in};n_{\omega\neq\omega_\mathrm{in}}=0\rangle_L \otimes |0\rangle_\mathrm{cav} \otimes |0\rangle_R
            \end{align}
            for $n_\mathrm{in}$ input photons at frequency $\omega_\mathrm{in}$ in the left-side free space modes. The cavity is in its dark ground state $|0\rangle_\mathrm{cav}$ which however does not necessarily have zero average photon number at ultrastrong coupling; the right-side bath modes and all other left-side bath modes are unpopulated.
            Using the Lippmann-Schwinger equation, or equivalently by expanding an adiabatic switch-on of photon tunneling $\hat{H}(t) = \hat{H}_0 + e^{\eta t} \hat{V}$, the final state can be related to the input state via \cite{bruus2004many}
            \begin{align}
            |\psi_i(t_f)\rangle = e^{-iE_i t_f}\bigg(|\phi_i\rangle + e^{\eta t_f}\hat{\mathcal{G}} \hat{T}|\phi_i\rangle\bigg)
            \end{align}
            where $\hat{\mathcal{G}} = (E_i-\hat{H}_0+i\eta)^{-1}$, the initial-state energy of $|\phi_i\rangle$ with respect to $\hat{H}_0$ is denoted as $E_i$, $\eta$ is an infinitesimal factor, and
            \begin{align}
            \hat{T} = \hat{V} \sum_{n=0}^\infty (\hat{\mathcal{G}}\hat{V})^n
            \end{align}
            We find that
            \begin{align}
            G^{(1)}(t_m) = \sum_{F} |\langle F| \hat{b}_{\mathrm{out},R}(t_m) e^{\eta t_f}\hat{\mathcal{G}}\hat{T}|\phi_i\rangle|^2
            \end{align}
            where we note that $\hat b_{\mathrm{out},R}(t_m)|\phi_i\rangle=0$ (the right-side bath is initially in its vacuum state). Substituting the definition for $\hat{b}_{\mathrm{out},R}(t)$ [Eq. (\ref{eq:bout})], we obtain
            \begin{align}
            G^{(1)} = \sum_{F} \bigg|\int_0^\infty d\omega \ e^{i \omega t_f} \langle F|\hat b_R(\omega) \hat{\mathcal{G}}\hat{T}|\phi_i\rangle\bigg|^2
            \end{align}
            where $t_m$ can be discarded in the limit $t_f \to +\infty$ while keeping $\eta t_f \to 0$.
            Next, $G^{(2)}$ at coincidence is
            \begin{align}
            G^{(2)}(t_m) = \langle \hat b_{\mathrm{out},R}^\dag(t_m) \hat b_{\mathrm{out},R}^\dag(t_m) \hat b_{\mathrm{out},R}(t_m) \hat b_{\mathrm{out},R}(t_m)\rangle
            \end{align}
            from which we obtain a similar expression
            \begin{align}
            G^{(2)} = \sum_F \left|\int_0^\infty d\omega \int_0^\infty d\omega'\ 2e^{i(\omega+\omega')t_f} \langle F|\hat{b}_R(\omega)\hat{b}_R(\omega')\hat{\mathcal{G}}\hat{T}|\phi_i\rangle\right|^2
            \end{align}
            We now collect all relevant scattering matrix elements (taking $\eta t_f\to 0$) by expanding $\hat{T}$ in powers of $\hat{V}$. Writing $G^{(2)} = \sum_F | \sum_n M_n |^2$, the matrix elements read
            \begin{align}
            M_n &= \int_0^\infty d\omega \int_0^\infty d\omega'\ 2e^{i(\omega+\omega')t_f} \langle F|\hat{b}_R(\omega)\hat{b}_R(\omega')\hat{\mathcal{G}} \hat{T}_n|\phi_i\rangle\\
            &= \int_0^\infty d\omega \int_0^\infty d\omega'\ \frac{2e^{i(\omega+\omega')t_f}}{n_\mathrm{in}\omega_\mathrm{in}-(E_f-E_0)-\omega-\omega'+i\eta}\, N_n(\omega,\omega')
            \end{align}
            with
            \begin{align}
            N_n(\omega,\omega') = \langle F|\hat{T}_n|0\rangle
            \end{align}
            These expressions enumerate different scattering processes where $n_\mathrm{in}$ input photons at frequency $\omega_\mathrm{in}$ tunnel through the cavity and precisely two photons moved to the right-side bath (detector) in the final state. Here, $T_n$ are individual terms in the scattering expansion, and $E_0$, $E_f$ are the intra-cavity ground state and final state energies, respectively.
            In principle, $T$ contains an infinitude of scattering processes; we truncate at order $\hat{V}^4$ ($\hat{V}^2$) for $G^{(2)}$ ($G^{(1)}$), which permits the tunneling of one or two photons from the input (left) to the output (right).

        \subsection{Contour integral identity}
    
            Evaluating each scattering matrix element in the non-Markovian input output formulation requires computing one-sided (positive) output-photon frequency integrals in the limit $t_f \to \infty$. These integrals take the form of
            \begin{align}
                \left. \int_{0}^{\infty} d\omega~ f(\omega)~ e^{i \omega t_f} \right|_{t_f \to \infty}
            \end{align}
            where $f(\omega)$ is a meromorphic function with poles in the upper half plane that vanishes for $|\omega| \to \infty$ for $0 \leq \arg \omega \leq \pi/2$.
            
            \begin{figure}[t]
                \centering
                \includegraphics[width=0.3\linewidth]{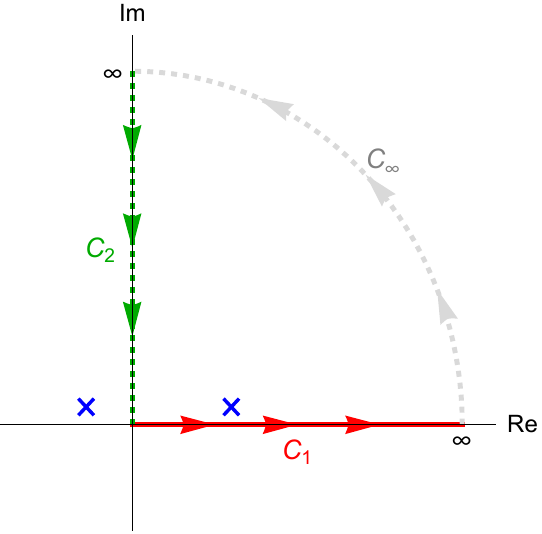}
                \caption{Contour of integration for frequency integrals in the limit $t_f \to \infty$. The red contour $C_1$ denotes the original one-sided frequency integrals in non-Markovian input-output theory. The gray dashed contour $C_\infty$ vanishes trivially. The green dashed contour $C_2$ vanishes in the limit $t_f \to \infty$, permitting a straightforward evaluation of $\int_{C_1}$ via summing the residues within the upper-right quadrant of the complex plane.}
                \label{fig:contour}
            \end{figure}
            
            In the limit $t_f \to \infty$, this integral can be computed straightforwardly by closing the integration contour in the upper-right quadrant of the complex plane, shown in Fig. \ref{fig:contour}. Here, $C_1$ is the desired integral and $C_{\infty}$ vanishes trivially. The integral along contour $C_2$ (the imaginary frequency axis) reads
            \begin{align}
                \left. \int_{C_2} dz~ f(z)~ e^{i z t_f} \right|_{t_f \to \infty} = \left. -\int_{0}^{\infty} ds~ f(is)~ e^{-st_f} \right|_{t_f \to \infty} \to 0
            \end{align}
            and vanishes in the limit $t_f \to \infty$. We therefore find that
            \begin{align}\label{eq:contour-integral}
                \left. \int_{0}^{\infty} d\omega~ f(\omega)~ e^{i \omega t_f} \right|_{t_f \to \infty} = 2\pi i \sum_{z_j}~ \theta(\mathrm{Re}\, z_j) \mathop{\mathrm{Res}}_{z \to z_j} f(z) ~e^{i z t_f}
            \end{align}
            Note that $\mathrm{Im}\, z_j = \eta$; hence, the residue remains finite in the limit $t_f \to \infty$, by ensuring that $\eta t_f \to 0$ as described above.

    \subsection{Single-photon detection $G^{(1)}$}

        For $G^{(1)}$, only three scattering processes $O$, $OI$, and $IO$ contribute at leading order. First, consider process $O$, the tunneling of a dark-cavity photon to the detector without an input field. We denote the intra-cavity ground state and final state as $|0\rangle$ and $|f\rangle$, respectively. The relevant tunneling matrix element can then be simplified as $\sqrt{\gamma/2\pi} \langle F | \hat{b}_R(\omega) \hat{a} | \phi_i \rangle = \sqrt{\gamma/2\pi} \langle f | \hat{a} | 0 \rangle$. Starting from the cavity ground state, one can evaluate the integral
        \begin{align}
        I_O = \int_0^\infty d\omega\ \sqrt{\frac{\gamma}{2\pi}} e^{i\omega t_f} \frac{\langle f|\hat a|0\rangle}{E_0-E_f -\omega + i\eta} = -2\pi i \sqrt{\frac{\gamma}{2\pi}} \langle f|\hat a|0\rangle\, \theta(E_0-E_f) e^{i(E_0-E_f+i\eta)t_f} =0
        \end{align}
        using Eq. \ref{eq:contour-integral}, and find that it vanishes identically as $E_f$ must be greater than $E_0$. Crucially this tells us that a dark cavity in its ground state cannot radiate. This result is expected, but not always reproduced by works in the literature.
        
        Next we have the $OI$ integral, which involves an intermediate state where one dark-cavity photon tunnels to the detector. Inserting a complete set of intermediate many-body states for the cavity, one obtains
        \begin{align}
        I_{OI}^{\,i} &= \frac{\gamma}{2\pi} \langle f|\hat a^\dag |i\rangle\langle i|\hat a|0\rangle \int_0^\infty d\omega\ \frac{e^{i\omega t_f}}{E_0-E_f +\omega_\mathrm{in} -\omega + i\eta} \frac{1}{E_0-E_i -\omega + i\eta}\\
        &= 2\pi i \frac{\gamma}{2\pi} \frac{\langle f|\hat a^\dag |i\rangle\langle i|\hat a|0\rangle}{E_f-E_i-\omega_\mathrm{in}}\bigg[e^{i(E_0-E_i)t_f}\theta(E_0-E_i) - e^{i(E_0-E_f+\omega_\mathrm{in})t_f} \theta(E_0-E_f+\omega_\mathrm{in})\bigg]\\
        &= - i \gamma \frac{\langle f|\hat a^\dag |i\rangle\langle i|\hat a|0\rangle}{E_f-E_i-\omega_\mathrm{in}} \theta(E_0-E_f+\omega_\mathrm{in}) e^{i(E_0-E_f+\omega_\mathrm{in})t_f}
        \end{align}
        where we used the fact that $\theta(E_0 - E_i) = 0$ ($i$ cannot be the ground state due to conserved photon number parity).
        Finally for the $IO$ integral
        \begin{align}
        I_{IO}^{\,i} &= \frac{\gamma}{2\pi} \frac{\langle f|\hat a |i\rangle\langle i|\hat a^\dag|0\rangle}{E_0-E_i +\omega_\mathrm{in}} \int_0^\infty d\omega\ \frac{e^{i\omega t_f}}{E_0-E_f +\omega_\mathrm{in} -\omega}\\
        &= -i\gamma \frac{\langle f|\hat a |i\rangle\langle i|\hat a^\dag|0\rangle}{E_0-E_i +\omega_\mathrm{in} + i\eta} \theta(E_0-E_f+\omega_\mathrm{in}) e^{i(E_0-E_f+\omega_\mathrm{in})t_f}
        \end{align}
        Combining these we find
        \begin{align}\label{eq:G1-square}
        G^{(1)} = \gamma^2 \sum_f \left|\sum_i \theta(E_0-E_f+\omega_\mathrm{in}) \bigg[\frac{\langle f|\hat a^\dag |i\rangle\langle i|\hat a|0\rangle}{E_f-E_i-\omega_\mathrm{in}} + \frac{\langle f|\hat a |i\rangle\langle i|\hat a^\dag|0\rangle}{E_0-E_i +\omega_\mathrm{in} + i\eta}\bigg]\right|^2
        \end{align}

    \subsection{Two-photon detection $G^{(2)}$}

        We now compute $G^{(2)}$ and truncate to processes with exactly two output photons and at most two input photons since processes with more input/output photons will be suppressed by an extra power of $\sqrt{\gamma}$, and more/fewer output photons will not result in any contribution to $G^{(2)}$.
        This leaves us with ten processes $OO$, $OOI$, $OIO$, $IOO$, $OOII$, $OIOI$, $OIIO$, $IOOI$, $IOIO$, $IIOO$, where starting in time order from the cavity ground state we proceed right to left; we illustrate these processes in Fig. \ref{fig:s7}. $I$ becomes $\hat a^\dag$ and is accompanied by a factor $\omega_\mathrm{in}$ in the subsequent energy denominator, $O$ becomes $\hat a$ and is accompanied by a factor $\omega$ (first photon emitted) and $\omega'$ (second photon emitted) in the subsequent denominator. Explicitly the scattering processes read
        \begin{align}
        N_{OO}(\omega,\omega') &= \big(\frac{\gamma}{2\pi}\big) \langle f| \hat a \frac{1}{E_0-\omega-\hat H_\mathrm{cav}+i\eta} \hat a |0\rangle\\
        N_{OOI}(\omega,\omega') &= \big(\frac{\gamma}{2\pi}\big)^{3/2} \langle f| \hat a^\dag \frac{1}{E_0-\omega-\omega'-\hat H_\mathrm{cav}+i\eta} \hat a \frac{1}{E_0-\omega-\hat H_\mathrm{cav}+i\eta} \hat a |0\rangle\\
        N_{OIO}(\omega,\omega') &= \big(\frac{\gamma}{2\pi}\big)^{3/2} \langle f| \hat a \frac{1}{E_0+\omega_\mathrm{in}-\omega-\hat H_\mathrm{cav}+i\eta} \hat a^\dag \frac{1}{E_0-\omega-\hat H_\mathrm{cav}+i\eta} \hat a |0\rangle\\
        N_{IOO}(\omega,\omega') &= \big(\frac{\gamma}{2\pi}\big)^{3/2} \langle f| \hat a \frac{1}{E_0+\omega_\mathrm{in}-\omega-\hat H_\mathrm{cav}+i\eta} \hat a \frac{1}{E_0+\omega_\mathrm{in}-\hat H_\mathrm{cav}+i\eta} \hat a^\dag |0\rangle\\
        N_{OOII}(\omega,\omega') &= \big(\frac{\gamma}{2\pi}\big)^2 \langle f| \hat a^\dag \frac{1}{E_0+\omega_\mathrm{in}-\omega-\omega'-\hat H_\mathrm{cav}+i\eta} \hat a^\dag \frac{1}{E_0-\omega-\omega'-\hat H_\mathrm{cav}+i\eta} \hat a \frac{1}{E_0-\omega-\hat H_\mathrm{cav}+i\eta} \hat a |0\rangle\\
        N_{OIOI}(\omega,\omega') &= \big(\frac{\gamma}{2\pi}\big)^2 \langle f| \hat a^\dag \frac{1}{E_0+\omega_\mathrm{in}-\omega-\omega'-\hat H_\mathrm{cav}+i\eta} \hat a \frac{1}{E_0+\omega_\mathrm{in}-\omega-\hat H_\mathrm{cav}+i\eta} \hat a^\dag \frac{1}{E_0-\omega-\hat H_\mathrm{cav}+i\eta} \hat a |0\rangle\\
        N_{OIIO}(\omega,\omega') &= \big(\frac{\gamma}{2\pi}\big)^2 \langle f| \hat a \frac{1}{E_0+2\omega_\mathrm{in}-\omega-\hat H_\mathrm{cav}+i\eta} \hat a^\dag \frac{1}{E_0+\omega_\mathrm{in}-\omega-\hat H_\mathrm{cav}+i\eta} \hat a^\dag \frac{1}{E_0-\omega-\hat H_\mathrm{cav}+i\eta} \hat a |0\rangle\\
        N_{IOOI}(\omega,\omega') &= \big(\frac{\gamma}{2\pi}\big)^2 \langle f| \hat a^\dag \frac{1}{E_0+\omega_\mathrm{in}-\omega-\omega'-\hat H_\mathrm{cav}+i\eta} \hat a \frac{1}{E_0+\omega_\mathrm{in}-\omega-\hat H_\mathrm{cav}+i\eta} \hat a \frac{1}{E_0+\omega_\mathrm{in}-\hat H_\mathrm{cav}+i\eta} \hat a^\dag |0\rangle\\
        N_{IOIO}(\omega,\omega') &= \big(\frac{\gamma}{2\pi}\big)^2 \langle f| \hat a \frac{1}{E_0+2\omega_\mathrm{in}-\omega-\hat H_\mathrm{cav}+i\eta} \hat a^\dag \frac{1}{E_0+\omega_\mathrm{in}-\omega-\hat H_\mathrm{cav}+i\eta} \hat a \frac{1}{E_0+\omega_\mathrm{in}-\hat H_\mathrm{cav}+i\eta} \hat a^\dag |0\rangle\\
        N_{IIOO}(\omega,\omega') &= \big(\frac{\gamma}{2\pi}\big)^2 \langle f| \hat a \frac{1}{E_0+2\omega_\mathrm{in}-\omega-\hat H_\mathrm{cav}+i\eta} \hat a \frac{1}{E_0+2\omega_\mathrm{in}-\hat H_\mathrm{cav}+i\eta} \hat a^\dag \frac{1}{E_0+\omega_\mathrm{in}-\hat H_\mathrm{cav}+i\eta} \hat a^\dag |0\rangle
        \end{align}

        \begin{figure}
            \centering
            \includegraphics[width=0.5\linewidth]{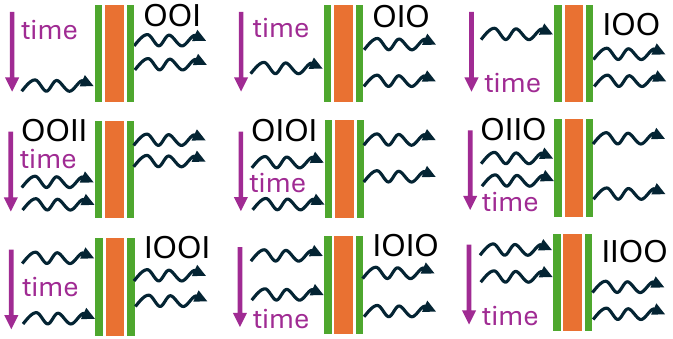}
            \caption{Scattering pathways at ultrastrong coupling, with exactly two output photons for up to two input photons. An $OO$ process vanishes for a cavity initially in its ground state.}
            \label{fig:s7}
        \end{figure}

    \subsubsection{Frequency Integrals}
    
        Now we will be interested in evaluating integrals over the frequencies of the two output photons
        \begin{align}
        M_n = \int_0^\infty d\omega \int_0^\infty d\omega'\ \frac{2e^{i(\omega+\omega')t_f}}{n_\mathrm{in}\omega_\mathrm{in}-(E_f-E_0)-\omega-\omega'+i\eta}\, N_n(\omega,\omega')
        \end{align}
        
        To evaluate these integrals we will insert a complete basis, and use the residue theorem to evaluate these integrals.

        \subsubsection{Example: $IIOO$ Process}
        
        Let us illustrate this for the $N_{IIOO}$ process and then quote the results for the other terms. Inserting the complete cavity basis three times, we have
        \begin{align}
        N_{IIOO}(\omega,\omega') &= \big(\frac{\gamma}{2\pi}\big)^2 \sum_{ijk} \langle f| \hat a|k\rangle\langle k|\hat a|j\rangle\langle j| \hat a^\dag |i\rangle\langle i|\hat a^\dag |0\rangle \notag\\ &\hspace{0.5 in}\times \frac{1}{E_0-E_k+2\omega_\mathrm{in}-\omega+i\eta}  \frac{1}{E_0-E_j+2\omega_\mathrm{in}+i\eta}  \frac{1}{E_0-E_i+\omega_\mathrm{in}+i\eta} 
        \end{align}
        For the $\omega'$ integral we have
        {\small\begin{align}
        &\int_0^\infty d\omega \int_0^\infty d\omega'\ \frac{2e^{i(\omega+\omega')t_f}}{2\omega_\mathrm{in}-(E_f-E_0)-\omega-\omega'+i\eta}\, \frac{1}{E_0-E_k+2\omega_\mathrm{in}-\omega+i\eta}  \frac{1}{E_0-E_j+2\omega_\mathrm{in}+i\eta}  \frac{1}{E_0-E_i+\omega_\mathrm{in}+i\eta}\\
        =& \frac{1}{E_0-E_j+2\omega_\mathrm{in}+i\eta}  \frac{1}{E_0-E_i+\omega_\mathrm{in}+i\eta} \int_0^\infty d\omega\ \frac{2 e^{i\omega t_f}}{E_0-E_k+2\omega_\mathrm{in}-\omega+i\eta} \int_0^\infty d\omega'\ \frac{e^{i\omega't_f}}{2\omega_\mathrm{in}-(E_f-E_0)-\omega-\omega'+i\eta}\\
        =& -4\pi i \frac{1}{E_0-E_j+2\omega_\mathrm{in}+i\eta}  \frac{1}{E_0-E_i+\omega_\mathrm{in}+i\eta} e^{it_f(2\omega_\mathrm{in}-(E_f-E_0)+i\eta)} \int_0^{2\omega_\mathrm{in}-(E_f-E_0)} d\omega\ \frac{\theta(E_0+2\omega_\mathrm{in}-E_f)}{E_0-E_k+2\omega_\mathrm{in}-\omega+i\eta}
        \end{align}}
        Where we use the identity in Eq. \ref{eq:contour-integral} to find
        \begin{align}
        &\int_0^\infty d\omega'\ \frac{e^{i\omega't_f}}{2\omega_\mathrm{in}-(E_f-E_0)-\omega-\omega'+i\eta}\\ &= 2\pi i \theta(2\omega_\mathrm{in}-(E_f-E_0)-\omega) \mathrm{Res}_{z\to 2\omega_\mathrm{in}-(E_f-E_0)-\omega+ i\eta} \frac{e^{izt_f}}{2\omega_\mathrm{in}-(E_f-E_0)-\omega-z+i\eta}\\
        &= -2\pi i \theta(2\omega_\mathrm{in}-(E_f-E_0)-\omega) e^{it_f(2\omega_\mathrm{in}-(E_f-E_0)-\omega+i\eta)}
        \end{align}
        
        The $\omega$ integral is then an integral over a finite frequency range and gives a log. Integrating,
        \begin{align}
        \int_0^{2\omega_\mathrm{in}-(E_f-E_0)} d\omega\ \frac{1}{E_0-E_k+2\omega_\mathrm{in}-\omega+i\eta} = \log(E_0-E_k+2\omega_\mathrm{in}+i\eta) - \log(E_f-E_k+i\eta)
        \end{align}
        recombining we find
        \begin{align}
        M_{IIOO} &= -4\pi i \big(\frac{\gamma}{2\pi}\big)^2 \sum_{ijk} \langle f| \hat a|k\rangle\langle k|\hat a|j\rangle\langle j| \hat a^\dag |i\rangle\langle i|\hat a^\dag |0\rangle\ e^{it_f(2\omega_\mathrm{in}-(E_f-E_0)+i\eta)} \theta(E_0+2\omega_\mathrm{in}-E_f)\notag\\&\hspace{1 in}\, \times\frac{1}{E_0-E_j+2\omega_\mathrm{in}+i\eta}  \frac{1}{E_0-E_i+\omega_\mathrm{in}+i\eta}  [\log(E_0-E_k+2\omega_\mathrm{in}+i\eta) - \log(E_f-E_k+i\eta)]
        \end{align}
        taking the $\eta t_f\to 0$ limit
        \begin{align}
        M_{IIOO} &= 4\pi i \big(\frac{\gamma}{2\pi}\big)^2 \sum_{ijk} \langle f| \hat a|k\rangle\langle k|\hat a|j\rangle\langle j| \hat a^\dag |i\rangle\langle i|\hat a^\dag |0\rangle  \theta(E_0+2\omega_\mathrm{in}-E_f) \notag\\&\hspace{1 in} \times \frac{\log(E_f-E_k+i\eta) - \log(E_0-E_k+2\omega_\mathrm{in}+i\eta)}{(E_0-E_j+2\omega_\mathrm{in}+i\eta)(E_0-E_i+\omega_\mathrm{in}+i\eta)}
        \end{align}

    \subsubsection{Scattering Matrix expressions for $G^{(2)}$}
        
        Similar techniques can be applied to the remaining scattering matrix elements. Here, we tabulate all relevant processes that are included in the calculations presented in the main text:
        \begin{align}
            M_{OO} &= 0\\
            M_{OOI} &= 4\pi i\big(\frac{\gamma}{2\pi}\big)^{3/2} \sum_{ij} \langle f| \hat a^\dag |j\rangle\langle j| \hat a |i\rangle\langle i| \hat a |0\rangle \theta(E_0+\omega_\mathrm{in}-E_f)\frac{\log(E_0-E_i+i\eta) -\log(E_f-E_i-\omega_\mathrm{in}+i\eta)}{E_j-E_f+\omega_\mathrm{in}} \\
            M_{OIO} &= 4\pi i\big(\frac{\gamma}{2\pi}\big)^{3/2} \sum_{ij} \langle f| \hat a |j\rangle\langle j| \hat a^\dag |i\rangle\langle i| \hat a |0\rangle \theta(E_0+\omega_\mathrm{in}-E_f)\notag\\&\hspace{0.5 in} \frac{\log(E_0-E_i+i\eta) + \log(E_f-E_j+i\eta) - \log(E_f-E_i-\omega_\mathrm{in}+i\eta) - \log(E_0-E_j+\omega_\mathrm{in}+i\eta)}{E_j-E_i-\omega_\mathrm{in}}\\
            M_{IOO} &= 4\pi i\big(\frac{\gamma}{2\pi}\big)^{3/2} \sum_{ij} \langle f| \hat a |j\rangle\langle j| \hat a |i\rangle\langle i| \hat a^\dag |0\rangle \theta(E_0+\omega_\mathrm{in}-E_f)\notag\\&\hspace{0.5 in} \frac{\log(E_f-E_j+i\eta) - \log(E_f-E_i+i\eta) - \log(E_0-E_j+\omega_\mathrm{in}+i\eta) + \log(E_0-E_i+\omega_\mathrm{in}+i\eta)}{E_j-E_i} \\
            M_{OOII} &= 4\pi i\big(\frac{\gamma}{2\pi}\big)^2 \sum_{ijk} \langle f| \hat a^\dag |k\rangle\langle k| \hat a^\dag |j\rangle\langle j| \hat a |i\rangle\langle i| \hat a |0\rangle \frac{1}{E_k-E_f+\omega_\mathrm{in}} \notag\\&\hspace{0.1 in} \bigg[-\theta(E_0+\omega_\mathrm{in}-E_k)\frac{\log(E_0-E_i+i\eta)-\log(E_k-E_i-\omega_\mathrm{in}+i\eta)}{E_k-E_j-\omega_\mathrm{in}}\notag\\&\hspace{0.18 in} - \theta(E_0+2\omega_\mathrm{in}-E_f)\frac{\log(E_0-E_i+i\eta) - \log(E_f-E_i-2\omega_\mathrm{in}+i\eta)}{E_j-E_f+2\omega_\mathrm{in}}\bigg]\\
            M_{OIOI} &= 4\pi i\big(\frac{\gamma}{2\pi}\big)^2 \sum_{ijk} \langle f| \hat a^\dag |k\rangle\langle k| \hat a |j\rangle\langle j| \hat a^\dag |i\rangle\langle i| \hat a |0\rangle \frac{1}{(E_k-E_f+\omega_\mathrm{in})(E_i-E_j+\omega_\mathrm{in})} \notag\\
            &\hspace{0.1 in} \bigg[-\theta(E_0+\omega_\mathrm{in}-E_k)[\log(E_k-E_j+i\eta) + \log(E_0-E_i+i\eta)\notag\\&\hspace{1.2 in} - \log(E_k-E_i-\omega_\mathrm{in}+i\eta) - \log(E_0-E_j+\omega_\mathrm{in}+i\eta)] \notag\\&\hspace{0.18 in} + \theta(E_0+2\omega_\mathrm{in}-E_f) [\log(E_0-E_i+i\eta)-\log(E_f-E_i-2\omega_\mathrm{in}+i\eta)\notag\\&\hspace{1.2 in}+\log(E_f-E_j-\omega_\mathrm{in}+i\eta)-\log(E_0-E_j+\omega_\mathrm{in}+i\eta)]\bigg]\\
            M_{OIIO} &= 4\pi i\big(\frac{\gamma}{2\pi}\big)^2 \sum_{ijk} \langle f| \hat a |k\rangle\langle k| \hat a^\dag |j\rangle\langle j| \hat a^\dag |i\rangle\langle i| \hat a |0\rangle \frac{1}{E_i-E_j+\omega_\mathrm{in}} \theta(E_0+2\omega_\mathrm{in}-E_f)\notag\\&\hspace{0.1 in} \bigg[\frac{\log(E_0-E_i+i\eta) + \log(E_f-E_k+i\eta) - \log(E_f-E_i-2\omega_\mathrm{in}+i\eta) - \log(E_0-E_k+2\omega_\mathrm{in} +i\eta)}{E_k-E_i-2\omega_\mathrm{in}}\notag\\&\hspace{0.1 in}  + \frac{-\log(E_f-E_k+i\eta) + \log(E_f-E_j-\omega_\mathrm{in}+i\eta) - \log(E_0-E_j+\omega_\mathrm{in}+i\eta) + \log(E_0-E_k+2\omega_\mathrm{in}+i\eta)}{E_k-E_j-\omega_\mathrm{in}}\bigg] \\
            M_{IOOI} &= 4\pi i\big(\frac{\gamma}{2\pi}\big)^2 \sum_{ijk} \langle f| \hat a^\dag |k\rangle\langle k| \hat a |j\rangle\langle j| \hat a |i\rangle\langle i| \hat a^\dag |0\rangle \bigg[
            \theta(E_0+\omega_\mathrm{in}-E_k)\frac{\log(E_k-E_j+i\eta) - \log(E_0-E_j+\omega_\mathrm{in}+i\eta)}{(E_k-E_f+\omega_\mathrm{in})(E_0-E_i+\omega_\mathrm{in}+i\eta)} \notag\\
            &\hspace{0.1 in} -\theta(E_0+2\omega_\mathrm{in}-E_f)\frac{\log(E_f-E_j-\omega_\mathrm{in}+i\eta) - \log(E_0-E_j+\omega_\mathrm{in}+i\eta)}{(E_k-E_f+\omega_\mathrm{in})(E_0-E_i+\omega_\mathrm{in}+i\eta)}\bigg]\\
            M_{IOIO} &= 4\pi i\big(\frac{\gamma}{2\pi}\big)^2 \sum_{ijk} \langle f| \hat a |k\rangle\langle k| \hat a^\dag |j\rangle\langle j| \hat a |i\rangle\langle i| \hat a^\dag |0\rangle\theta(E_0+2\omega_\mathrm{in}-E_f)\notag\\&\hspace{0.1 in}  \frac{\log(E_f-E_k+i\eta)-\log(E_f-E_j-\omega_\mathrm{in}+i\eta) + \log(E_0-E_j+\omega_\mathrm{in}+i\eta) - \log(E_0-E_k+2\omega_\mathrm{in}+i\eta)}{(E_k-E_j-\omega_\mathrm{in})(E_0-E_i+\omega_\mathrm{in}+i\eta)} \label{eq:M_IOIO} \\
            M_{IIOO} &= 4\pi i\big(\frac{\gamma}{2\pi}\big)^2 \sum_{ijk} \langle f| \hat a |k\rangle\langle k| \hat a |j\rangle\langle j| \hat a^\dag |i\rangle\langle i| \hat a^\dag |0\rangle \, \theta(E_0+2\omega_\mathrm{in}-E_f) \frac{\log(E_f-E_k+i\eta)-\log(E_0-E_k+2\omega_\mathrm{in}+i\eta)}{(E_0-E_j+2\omega_\mathrm{in}+i\eta)(E_0-E_i+\omega_\mathrm{in}+i\eta)} \label{eq:M_IIOO}
        \end{align}
        Combining all processes, we finally obtain $G^{(2)}$ for pairs of output photons at coincidence, as a sum over all matrix elements
        \begin{align}
            G^{(2)} = \sum_f \bigg|\sum_n M_n\bigg|^2
        \end{align}
        where $n$ runs over 9 processes, and $f$ runs over all possible final cavity-material states, which can be computed using the exact diagonalization of $\hat{H}_\mathrm{cav}$.

    \subsection{Recovery of RWA Lindbladian Limit}\label{sec:lindblad-limit}

        We begin by noting that in the RWA only one process survives in $G^{(1)}$ from Eq. \ref{eq:G1-square}
        \[
        G^{(1)}(0) &= \gamma^2 \sum_f \left|\sum_i \theta(E_0-E_f+\omega_\mathrm{in}) \bigg[\frac{\langle f|\hat a^\dag |i\rangle\langle i|\hat a|0\rangle}{E_f-E_i-\omega_\mathrm{in}} + \frac{\langle f|\hat a |i\rangle\langle i|\hat a^\dag|0\rangle}{E_0-E_i +\omega_\mathrm{in} + i\eta}\bigg]\right|^2\\
        &= \gamma^2 \left|\frac{\langle 0,0|\hat a |1,0\rangle\langle 1,0|\hat a^\dag|0,0\rangle}{E_{0,0}-E_{1,0} +\omega_\mathrm{in} + i\eta}\right|^2\\
        &= \gamma^2 \left|\frac{1}{-\omega_\mathrm{cav}+i\gamma+\omega_\mathrm{in} + i\eta}\right|^2\\
        &= \frac{\gamma^2}{(\omega_\mathrm{in}-\omega_\mathrm{cav})^2 + \gamma^2}\\
        &= \frac{\gamma^2}{\delta_\mathrm{cav}^2 + \gamma^2}
        \]
        where $\gamma$ in the denominator emerges from the non-Hermitian broadening and we set $\eta\to 0^+$.

        Naively, two processes survive in $G^{(2)}(0)$. First
        \[
        M_{IOIO} &= 4\pi i\big(\frac{\gamma}{2\pi}\big)^2 \langle 0,0| \hat a |1,0\rangle\langle 1,0| \hat a^\dag |0,0\rangle\langle 0,0| \hat a |1,0\rangle\langle 1,0| \hat a^\dag |0,0\rangle\theta(2\omega_\mathrm{in})\notag\\&\hspace{0.1 in}  \frac{\log(E_{0,0}-E_{1,0}+i\eta)-\log(-\omega_\mathrm{in}+i\eta) + \log(\omega_\mathrm{in}+i\eta) - \log(E_{0,0}-E_{1,0}+2\omega_\mathrm{in}+i\eta)}{(E_{1,0}-E_{0,0}-\omega_\mathrm{in})(E_{0,0}-E_{1,0}+\omega_\mathrm{in}+i\eta)}\\
        &= 4\pi i\big(\frac{\gamma}{2\pi}\big)^2 \frac{\log(-\omega_\mathrm{cav}+i\eta)-\log(-\omega_\mathrm{in}+i\eta) + \log(\omega_\mathrm{in}+i\eta) - \log(-\omega_\mathrm{cav}+2\omega_\mathrm{in}+i\eta)}{(\omega_\mathrm{cav}-i\gamma-\omega_\mathrm{in})(-\omega_\mathrm{cav}+\omega_\mathrm{in}+i\gamma+i\eta)}\\
        &= \frac{i\gamma^2}{\pi} \frac{\log(\omega_\mathrm{cav}/(2\omega_\mathrm{in}-\omega_\mathrm{cav}))}{(\delta_\mathrm{cav}-i\gamma)(-\delta_\mathrm{cav}+i\gamma)}\\
        &= 0
        \]
        where we took Eq. \ref{eq:M_IOIO} and noted that $|0\rangle=|0,0\rangle$ and $|i\rangle=|k\rangle=|1,0\rangle$ with $|j\rangle=|0,0\rangle$ for $IOIO$. In the last line we took $\delta_\mathrm{cav}/\omega_\mathrm{cav}\to 0$ as this recovers the Markovian limit, where the logarithms that emerge from retaining only positive bath frequencies vanish when $\delta_\mathrm{cav}/\omega_\mathrm{cav}\to 0$.

        Second we consider
        \[
        M_{IIOO} &= 4\pi i\big(\frac{\gamma}{2\pi}\big)^2 \sum_{j} \langle 0,0| \hat a |1,0\rangle\langle 1,0| \hat a |j\rangle\langle j| \hat a^\dag |1,0\rangle\langle 1,0| \hat a^\dag |0,0\rangle \, \theta(2\omega_\mathrm{in}) \frac{\log(E_{0,0}-E_{1,0}+i\eta)-\log(E_{0,0}-E_{1,0}+2\omega_\mathrm{in}+i\eta)}{(E_{0,0}-E_j+2\omega_\mathrm{in}+i\eta)(E_{0,0}-E_{1,0}+\omega_\mathrm{in}+i\eta)}\\
        &= 4\pi i\big(\frac{\gamma}{2\pi}\big)^2 \sum_{j} \langle 1,0| \hat a |j\rangle\langle j| \hat a^\dag |1,0\rangle \, \frac{i\pi}{(E_{0,0}-E_j+2\omega_\mathrm{in}+i\eta)(-\omega_\mathrm{cav}+i\gamma+\omega_\mathrm{in}+i\eta)}\\
        &= \frac{2\gamma^2}{(\delta_\mathrm{cav}-i\gamma)}  \langle 2,0| \bigg[\sum_{j} \frac{|j\rangle\langle j|}{2\omega_\mathrm{in}-E_j}\bigg] |2,0\rangle
        \]
        where we took Eq. \ref{eq:M_IIOO} and noted that $|0\rangle=|0,0\rangle$ and $|i\rangle=|k\rangle=|1,0\rangle$ for $IIOO$. In the second to last line we again took the Markovian limit where the real logarithms vanish, and took $\eta\to 0^+$. Note that in RWA $E_{0,0}=0$.
        To proceed, we recall Eq. \ref{eq:polariton} where we now have the matrix element of the resolvent
        \[
        \bigg\langle 2,0\bigg| \begin{pmatrix}2\omega_\mathrm{in} - 2\omega_\mathrm{cav} + 2i\gamma & -\sqrt{2}\kappa \\ -\sqrt{2}\kappa & 2\omega_\mathrm{in} - \omega_h + i\gamma_h/2\end{pmatrix}^{-1} \bigg|2,0\bigg\rangle
        = \frac{2\omega_\mathrm{in} - \omega_h + i\gamma_h/2}{(2\omega_\mathrm{in} - 2\omega_\mathrm{cav} + 2i\gamma)(2\omega_\mathrm{in} - \omega_h + i\gamma_h/2) - 2\kappa^2}
        \]
        So that simplifying and evaluating
        \[
        G^{(2)}(0) &= |M_{IIOO}|^2\\
        &= \bigg[\frac{\gamma^2}{\delta_\mathrm{cav}^2+\gamma^2}\bigg]^2 \bigg|\frac{(\delta_\mathrm{cav}+i\gamma)(2\omega_\mathrm{in} - \omega_h + i\gamma_h/2)}{(\omega_\mathrm{in} - \omega_\mathrm{cav} + i\gamma)(2\omega_\mathrm{in} - \omega_h + i\gamma_h/2) - \kappa^2}\bigg|^2\\
        &= \bigg[\frac{\gamma^2}{\delta_\mathrm{cav}^2+\gamma^2}\bigg]^2 \bigg|\frac{(\delta_\mathrm{cav}+i\gamma)(-\delta_h + i\gamma_h/2)}{(-\delta_\mathrm{cav} + i\gamma)(-\delta_h + i\gamma_h/2) - \kappa^2}\bigg|^2
        \]
        where we see that the leading term cancels with $[G^{(1)}(0)]^2$ to yield upon simplification
        \begin{align}
            g^{(2)}(0) = \left|\frac{(\gamma+i\delta_\mathrm{cav})(\gamma_h/2+i\delta_h)}{\kappa^2 + (\gamma+i\delta_\mathrm{cav})(\gamma_h/2+i\delta_h)}\right|^2
        \end{align}
        which is Eq. \ref{eq:g2-resonance} with detunings $\delta_\mathrm{cav}=\omega_\mathrm{cav}-\omega_\mathrm{in}$ and $\delta_h=\omega_h-2\omega_\mathrm{in}$.

    \subsection{Non-Hermitian Broadening}

        The above derivation assumed that $\hat H_\mathrm{cav}$ is Hermitian and the only pathway for energy to enter or leave the cavity was through the cavity mirrors in powers of $\gamma$. Formally, a finite lifetime emerges from resumming an infinite series of processes in the expansion of $\hat{T}$, which introduces a finite lifetime to intermediate-state self energies. Here, we model the effects of finite cavity linewidth or material decoherence by introducing an effective non-Hermitian term to the cavity Hamiltonian. To do so, we replace the Hamiltonian with $\hat H_\mathrm{eff}=\hat H_\mathrm{cav} + \hat \Sigma$ where $\hat \Sigma = -i\gamma_\mathrm{cav} \hat{a}^\dag \hat{a} - i \gamma_h \hat{h}^\dag \hat{h}/2$. As this effective Hamiltonian is non-Hermitian, the many-body eigenstate decomposition of the cavity $|i\rangle \langle i|$ must be formulated in terms of the right and left eigenstates $|i^R\rangle\langle i^L|$. All scattering matrix element denominators hence become broadened by the photon number. To obtain a well-defined scattering state we insist that the log terms in the scattering matrix elements be interpreted as $\log(f(E)+i\eta)=\log(\mathrm{Re}[f(E)]+i\eta)$, forcing the branch points into the upper half plane. This is a natural extension to phenomenological broadening where the broadening enters in the denominators but not in the log to avoid accidental branch crossings. 
        We consider $\hat \Sigma=-i\gamma_\mathrm{cav}\hat a^\dag \hat a -i \gamma_h \hat h^\dag \hat h/2$ with $\gamma_\mathrm{cav}=\gamma$, generically $\gamma_\mathrm{cav}\geq\gamma$.

    \begin{figure*}[!b]
            \centering
            \includegraphics[width=\linewidth]{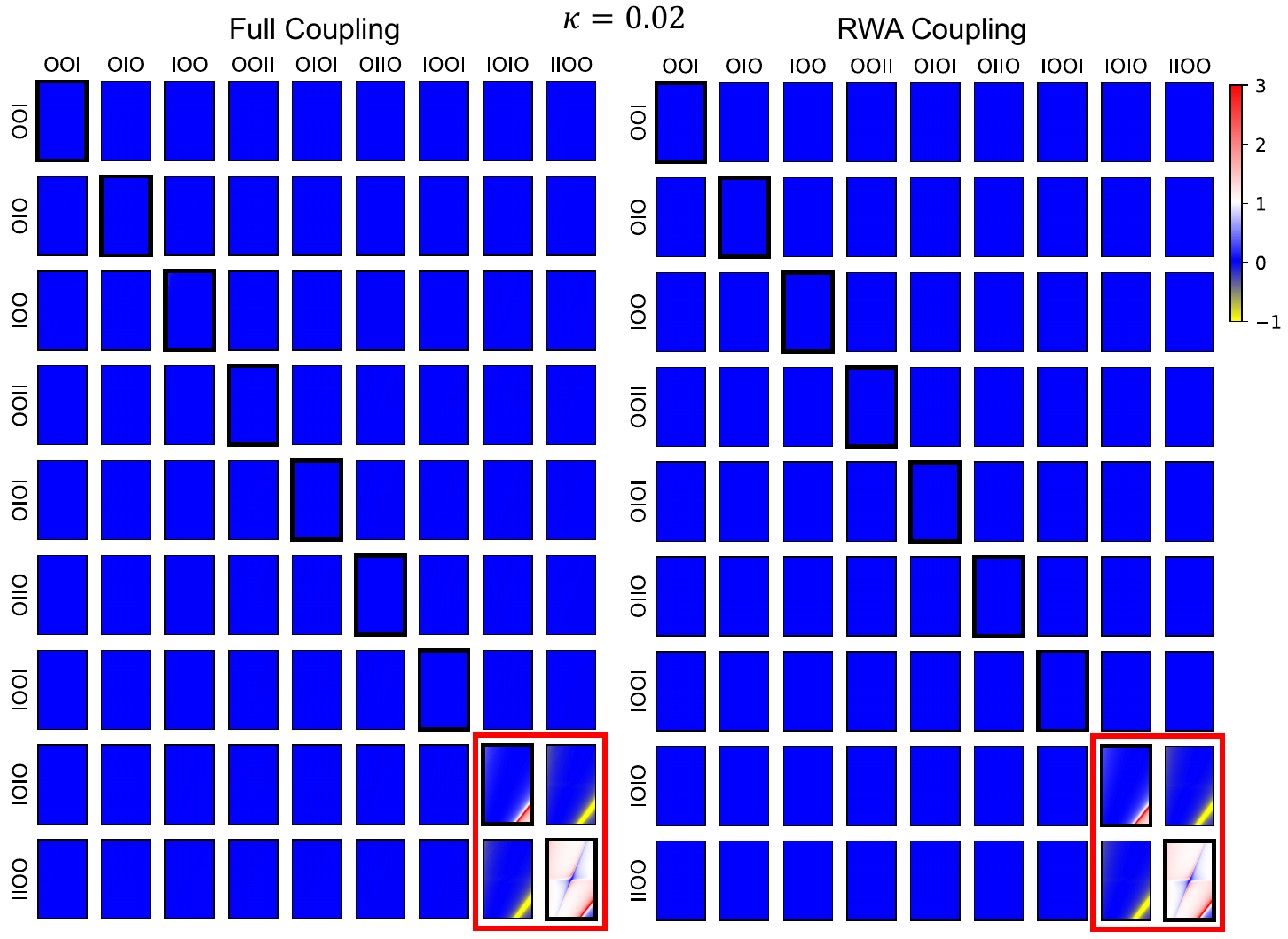}
            \caption{\textbf{(left)} Decomposition of $g^{(2)}(0)$ for Fig. \ref{fig:fig2}(a), $\kappa=0.02$, into constituent scattering pathways. Note that the diagonal entries are all non-negative, but that the off-diagonal terms can be negative (destructively interfering). The majority of terms make negligible contributions which is a reflection of the discrete level spectrum of the cavity-material system and the fact that even at ultrastrong coupling some terms are still strongly off-resonant. The processes boxed in red are the only processes that would be present in the RWA case and their sum is precisely what we plot in Fig. \ref{fig:fig2}(a). \textbf{(right)} Same as (left), but with the RWA coupling $\hat h^\dag \hat a\hat a+\hat h\hat a^\dag \hat a^\dag$ and no counter-rotating couplings. Note how all terms outside the red box vanish. Any new terms in Fig. \ref{fig:s8} can then be tied to the counter-rotating terms.}
            \label{fig:s8}
        \end{figure*}

    \section{Decomposition of Scattering Pathways}\label{sec:9x9}

        The scattering matrix approach to calculating $g^{(2)}$ involves a sum over nine two-photon emission processes, normalized via $[G^{(1)}]^2$. To gain further insight into individual contributions to $g^{(2)}$, Fig. \ref{fig:s8} shows all $9\times 9$ combinations of scattering processes that contribute to Fig. \ref{fig:fig2}(a), $\kappa=0.02$, by expanding $| \sum_n M_n |^2$. We show similar decompositions for Fig. \ref{fig:fig2}(b), $\kappa=0.05$, in Fig. \ref{fig:s9} and for Fig. Fig. \ref{fig:fig2}(c) in Fig. \ref{fig:s10}, $\kappa=0.10$. We see that at $\kappa=0.02$ in Fig. \ref{fig:s8} only the RWA scattering pathways boxed in red and and a little of the stimulated emission feature are visible. At $\kappa=0.05$ in Fig. \ref{fig:s9} we see that more scattering pathways have turned on including higher polariton features, and the stimulated emission feature has grown. In Fig. \ref{fig:s10}, at $\kappa=0.10$, we see that many additional features turn on; these features are largely obscured behind the stimulated emission peak, but some of these features might be able to be resolved in higher quality factor cavities. Finally, we decompose the individual features at $\kappa=0.10$ into their final states in Fig. \ref{fig:s11} where we see that for processes with an odd number of photons the cavity ends up in and odd parity state at the end (necessarily not the ground state); for processes with an even number of photons the cavity ends in an even parity state which can either be the ground state, or an excited state (as we see for the higher polariton features).

        \begin{figure*}
            \centering
            \includegraphics[width=\linewidth]{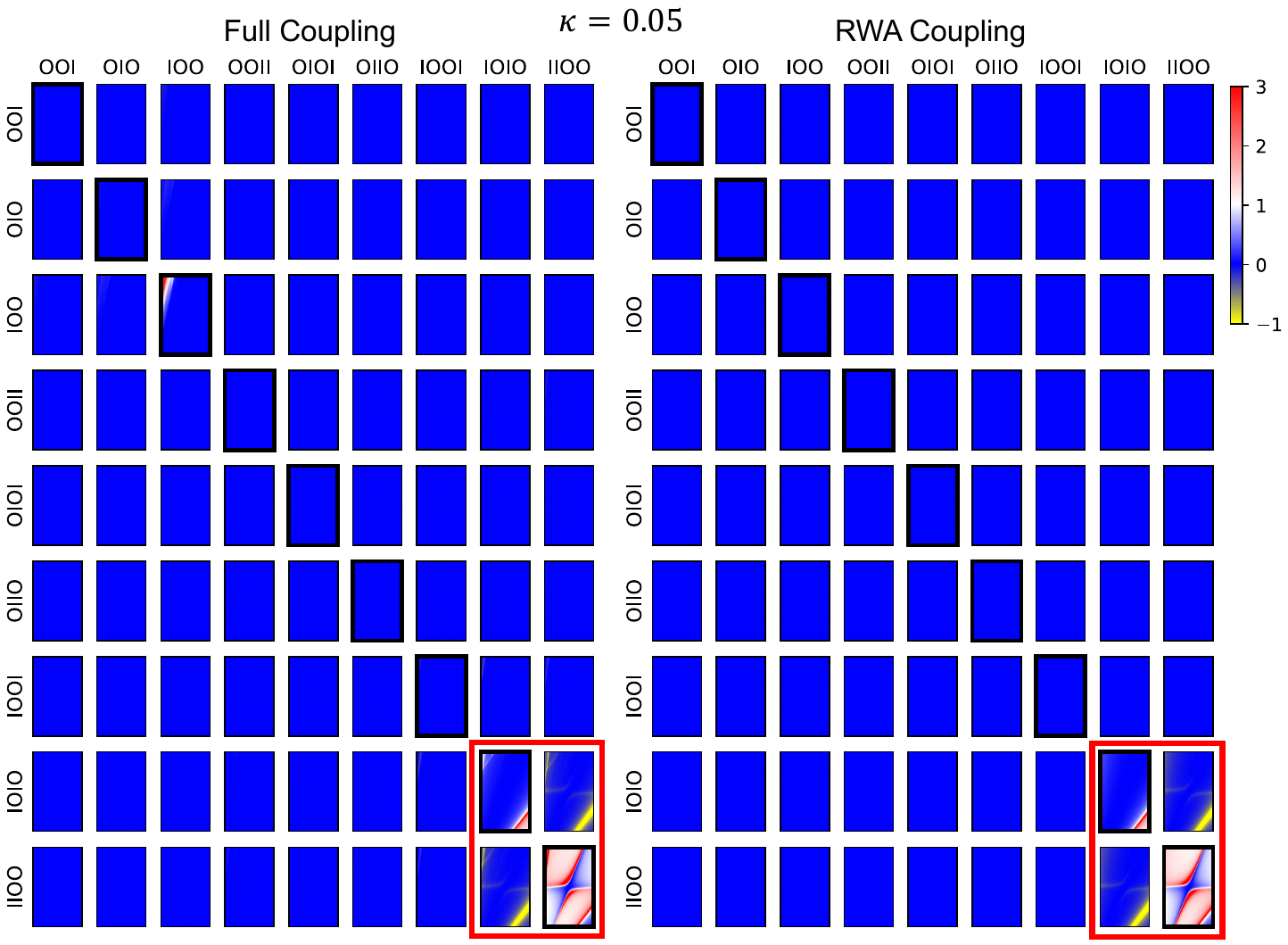}
            \caption{Same as \ref{fig:s8}, but for Fig. \ref{fig:fig2}(b), $\kappa=0.05$. Note the additional lines in the $IOIO$-$IOIO$ process, which come from the fact that there is vacuum occupation and transitions to states like $|3,0\rangle$ are possible with two input photons. Of note, the interference of the two processes can give negative contributions, and specifically the $IOIO$-$IIOO$ interference term cancels out a peak at low energy that would spuriously show up if we only considered the diagonal processes.}
            \label{fig:s9}
        \end{figure*}

        \begin{figure*}[!t]
            \centering
            \includegraphics[width=\linewidth]{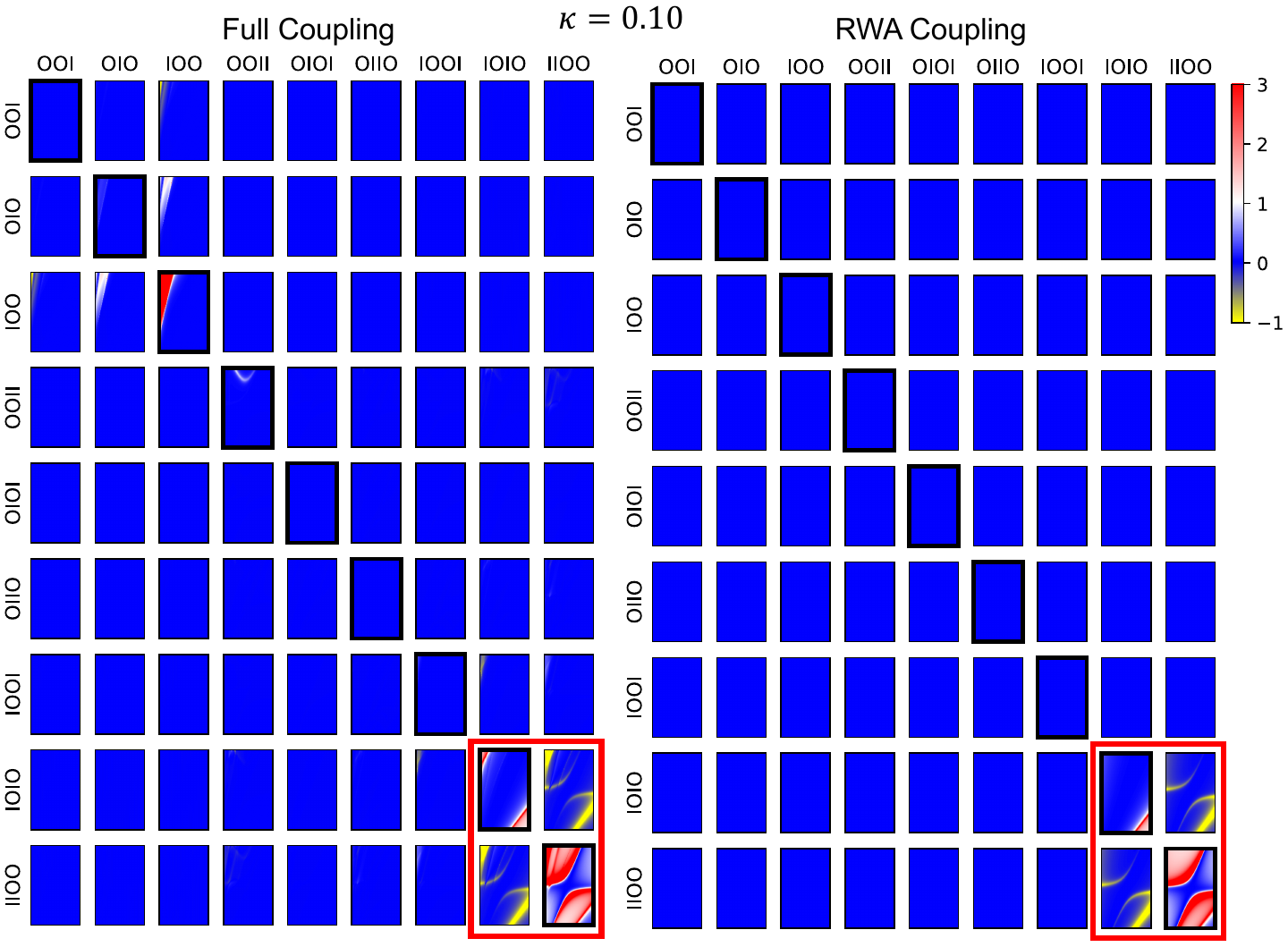}
            \caption{Same as \ref{fig:s8}, but for Fig. \ref{fig:fig2}(c), $\kappa=0.10$. Note the emergence of more scattering pathways, but most of these pathways are overshadowed by the $IOO$-$IOO$ process. One remaining process that is visible is the higher polariton in the $OOII$-$OOII$, which is analyzed in more detail in Fig. \ref{fig:s11}.}
            \label{fig:s10}
        \end{figure*}

        \begin{figure*}
            \centering
            \includegraphics[width=\linewidth]{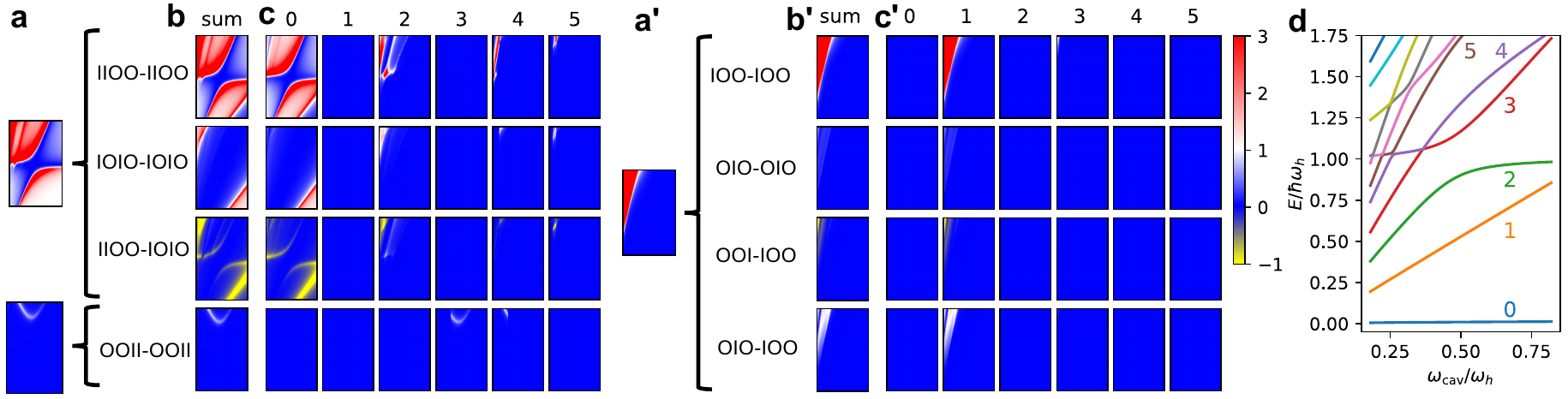}
            \caption{Further decomposition of Fig. \ref{fig:fig2}. \textbf{(a)} and \textbf{(a')} Sums of processes that produce the top two panels of Fig. \ref{fig:fig2}(c) with color scale extended to $-1$. The bottom panel ``all other processes" in the main text is not reproduced here, and instead we just plot the $OOII$-$OOII$ scattering process. \textbf{(b)} and \textbf{(b')} constituent scattering processes with contributions that are visible to the eye. Note that the off-diagonal terms are symmetric about the diagonal and we only plot one of the pair--i.e. $IOIO$-$IIOO$ has the same contribution as $IIOO$-$IOIO$. \textbf{(c)} and \textbf{(c')} Final state analysis. Note that all processes involving three photons must end in a parity odd state, and we see that they do as they end in the 1 state. The processes involving four photons must end in a parity even state such as 0, 2, or 4. The small bunching regions visible in final state 5 are there because numbering here is done by energy, not by eigenstate/parity. \textbf{(d)} Energy level scheme as a function of cavity frequency.}
            \label{fig:s11}
        \end{figure*}

        \phantom{.}

        \end{widetext}

        \phantom{.}

\end{document}